\documentclass[prd,preprint,eqsecnum,nofootinbib,amsmath,amssymb,tightenlines,dvips]{revtex4}
\usepackage{epsfig}
\usepackage{amssymb,latexsym,amsmath}
\newcommand{\eq}[1]{\begin{align} #1 \end{align}}

\newcommand{\p}{\partial}
\newcommand{\abs}[1]{\left\vert#1\right\vert}
\newcommand{\mc}[1]{\mathcal{#1}}

\begin{document}
\title{Towards precise calculation of transport coefficients in
the hadron gas. The shear and the bulk viscosities.}

\author{Oleg N. Moroz}
\affiliation{Bogolyubov Institute for Theoretical Physics, Kiev,
 Ukraine}

\email{ moroz@mail.bitp.kiev.ua }

\date {\today}

\begin{abstract}
The shear and the bulk viscosities of the hadron gas at low
temperatures are studied in the model with constant elastic cross
sections being relativistic generalization of the hard spheres
model. One effective radius ${r=0.4~fm}$ is chosen for all elastic
collisions. Only elastic collisions are considered which are
supposed to be dominant at temperatures ${T\leq 120-140~MeV}$. The
calculations are done in the framework of the Boltzmann equation
with the Boltzmann statistics distribution functions and the ideal
gas equation of state. The applicability of these approximations
is discussed. It's found that the bulk viscosity of the hadron gas
is much larger than the bulk viscosity of the pion gas while the
shear viscosity is found to be less sensitive to the mass spectrum
of hadrons. The constant cross sections and the Boltzmann
statistics approximation allows one not only to conduct precise
numerical calculations of transport coefficients in the hadron gas
but also to obtain some relatively simple relativistic analytical
closed-form expressions. Namely, the correct single-component
first-order shear viscosity coefficient is found. The
single-component first-order nonequilibrium distribution function,
some analytical results for the binary mixture and expressions for
mean collision rates, mean free paths and times are presented.
Comparison with some previous calculations for the hadron gas and
the pion gas is done too. This paper is the first step towards
calculations with inelastic processes included.
\end{abstract}



\maketitle

\section{ Introduction }
The bulk and the shear viscosity coefficients are transport
coefficients which enter in hydrodynamical equations and thus are
important for studying of nonequilibrium evolution of any
thermodynamic system.

Although experimental value of the shear viscosity was found to be
small in the hadron gas \cite{smalleta}, \cite{lacey} there are
two more additional reasons to study the shear viscosity. The
first one is the experimentally observed minimum of the ratio of
the shear viscosity to the entropy density $\eta/s$ near the
liquid-gas phase transition for different substances which may
help in studying of the quantum chromodynamics phase diagram and
finding of the location of the critical point
\cite{Csernai:2006zz}, \cite{lacey}. For a counterexample see
\cite{arXiv:1010.3119} and references therein. The second reason
is calculation of the $\eta/s$ in strongly interacting systems,
preferably real ones, for testing of the conjectured lowest bound
$\eta/s=\frac1{4\pi}$ \cite{adscftbound}, found in some conformal
field theories having dual gravity theories, and search of a new
one. The bound $\eta/s=\frac1{4\pi}$ was violated with different
counterexamples. For some reasonable ones see
\cite{Buchel:2008vz}, \cite{Sinha:2009ev}. Also see the recent
review \cite{arXiv:1108.0677}. The bulk viscosity may be not
negligibly small and is expected to have a maximum near the phase
transition \cite{Kharzeev:2007wb}, \cite{Karsch:2007jc},
\cite{chakkap}.

Whether one uses the Kubo formula or the Boltzmann equation one
faces nearly the same integral equation for transport coefficients
\cite{jeon}, \cite{jeonyaffe}, \cite{Arnold:2002zm}. The
preferable way to solve it is the variational (or Ritz) method.
Due to its complexity often the relaxation time approximation is
used in the framework of the Boltzmann equation. Though this
approximation is inaccurate, does not allow to control precision
of approximation and can potentially lead to large deviations. The
main difficulty in the variational method is in calculation of
collision integrals. To calculate any transport coefficient in the
lowest approximation in a mixture like the hadron gas with very
large number of components $N'$ one would need to calculate
roughly ${N'}^2$ special 12-dimensional integrals if only elastic
collisions are considered. Fortunately it's possible to simplify
these integrals considerably and perform these calculations in
reasonable time.

This paper contains calculations of the shear and the bulk
viscosity coefficients for the hadron-resonance gas in the
constant elastic differential cross sections model. The
calculations are done in the framework of the Boltzmann equation
with classical Boltzmann statistics, without medium effects and
with the ideal gas equation of state. The constant cross sections
and the Boltzmann statistics approximation allows one to obtain
some relatively simple analytical closed-form expressions.

The numerical calculations for the hadron-resonance gas presented
in this paper are the first step towards calculations with
number-changing (inelastic) processes taken into account. Then,
possibly, more realistic model of strong interactions and quantum
statistics will be implemented. These calculations can be
considered as quite precise at low temperatures where elastic
collisions dominate and equation of state is close to the ideal
gas equation of state.

The main applications of the results of this paper are to hydrodynamical description of heavy ion collisions,
like in \cite{Fogaca:2013cma}. See also the review of such applications \cite{Kapusta:2008vb}. Though,
analytical results of the present paper can be applied to other fields of physics, like cosmology,
see, e. g., \cite{BasteroGil:2012zr}. There are also other ways of dissipative description, which can
be mentioned, see, e. g. \cite{dissipcoef}.

\begin{figure}[h!]
\begin{center}
\epsfig{file=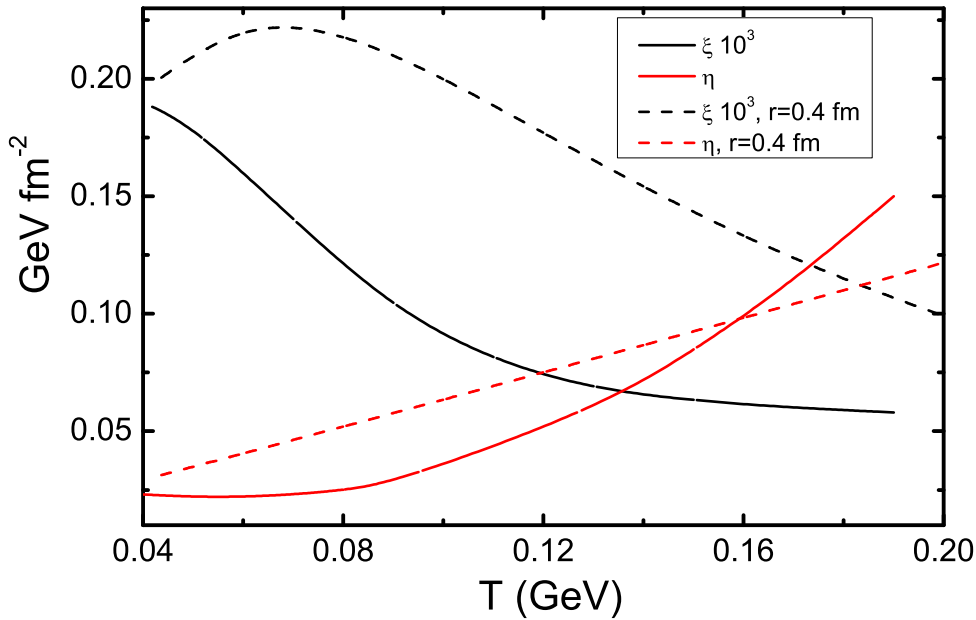,width=13cm} \caption{Dashed lines
designate calculations for the pion gas performed in the constant
cross sections approximation with the hard core radius $r=0.4~fm$.
The solid black line designates the 3-rd order bulk viscosity and
the solid red line designates the 1-st order shear viscosity of
the pion gas from \cite{prakash}, where the quasielastic (with
contribution of the $\rho$-resonance) semi-empirical $\pi\pi$
cross section was used. \label{CompViscosPions} }
\end{center}
\end{figure}

For comparison calculations for the pion gas are performed and the
results are found to be close to the results in \cite{prakash}
where only elastic processes were considered and the ideal gas
equation of state was used or to the results in \cite{davesne} (at
zero chemical potential) where the Bose statistics was used
instead of the Boltzmann one. The comparison is shown in fig.
\ref{CompViscosPions}. The discrepancies arise because of the
constant cross sections approximation used in this paper. Smaller
values of the viscosities in \cite{prakash} for some temperature
range originate because of the $\rho$-resonance contribution to
the quasielastic $\pi\pi$ cross section and the maximum of the
bulk viscosity is expected to be shifted towards lower
temperatures. In these calculations the nonvanishing value of the
bulk viscosity is obtained only due to the mass of
pions.\footnote{The bulk viscosity vanishes identically in scale
invariant theories \cite{weinberg}. Saying more concretely the
bulk viscosity vanishes identically if the distribution functions
are scale invariant (in the framework and approximation of the
common Boltzmann equation) \cite{jeonyaffe} which explains the
case of the massive monoatomic gas in the nonrelativistic theory
\cite{landau10} (where ${\epsilon-3P\neq0}$) with the scale
covariant energy spectrum $\frac{|\vec p|^2}{2m}$.}

In several papers the bulk viscosity $\xi$ was calculated for the
pion gas using the chiral perturbation theory and some other
approaches with quite large discrepancies between quantitative
results. In \cite{fernnicola} calculations were done by the Kubo
formula in rough approximation. There the number-changing
${2\leftrightarrow4}$ processes were neglected too and the
non-vanishing value of the bulk viscosity is obtained due to trace
anomaly and the mass of pions. At small temperatures where effects
of trace anomaly are small the magnitude of the bulk viscosity is
large in compare to the results of this paper and \cite{prakash},
\cite{davesne}. For example, at ${T=60~MeV}$ the bulk viscosity is
roughly 20 times more. The calculations in \cite{lumoore} are done
in the framework of the Boltzmann equation and have divergent
dependence of the $\xi(T)$ for ${T\rightarrow0}$ because of weak
${2\leftrightarrow4}$ number-changing processes taken into
account\footnote{This dependence should change at low
temperatures, where number-changing processes are expected to be
suppressed. See Sec. \ref{condappl} for more details.} and at
${T=140~MeV}$ the bulk viscosity is nearly 44 times larger than
the bulk viscosity calculated in this paper. Though calculations
in \cite{lumoore} were done in the first order approximation which
does not even take into account elastic collisions so that this
discrepancy can become smaller few times after more accurate
calculations. The function $\xi(T)$ may turn out to be not smooth
at the middle temperatures and the smooth function $\xi(T)$ is to
be obtained through extrapolation. In \cite{dobado} the bulk
viscosity was calculated in the framework of the Boltzmann
equation with the ideal gas equation of state and only elastic
collisions taken into account. The Inverse Amplitude Method was
used to get the scattering amplitudes of pions. The quantitative
results are close to the results in \cite{prakash, davesne}.
Calculations in the paper \cite{chenwang} are done in the
framework of the Boltzmann equation for massless pions. There the
bulk viscosity increases rapidly so that the ratio $\xi/s$
increases with the temperature.

Calculation of the shear viscosity in the hadron gas, done in
\cite{Gorenstein:2007mw} using some approximate formula and in
\cite{toneev} using relaxation time approximation, are in good
agreement with calculations in this paper\footnote{This is in
agreement with some tests of the relaxation time approximation \cite{RTA}}.
The bulk viscosity of
the hadron gas, calculated in \cite{toneev} has similar dependence
on the temperature though it is roughly 10-30 times more than the
bulk viscosity calculated in this paper. At low temperature
${T=100~MeV}$ and vanishing chemical potentials it is 20 times
more. In \cite{nhngr} calculation of the bulk viscosity was done
for the hadron-resonance gas with masses less than $2~GeV$ using
some special formula obtained though an ansatz for the small
frequency limit of the spectral density at zero spatial momentum
\cite{Kharzeev:2007wb}. There the bulk viscosity to entropy
density ratio $\xi/s$ has similar behavior to the $\xi/s$ in this
paper and is roughly 2 times more.

The bulk and the shear viscosities were calculated in the linear
$\sigma$-model with number-changing processes taken into account
and nonideal gas equation of state \cite{chakkap}. This model has
chiral symmetry restoration phase transition and can qualitatively
describe pion gas. These calculations demonstrate example that the
ratio of the shear viscosity to the entropy density $\eta/s$ has
the minimum near the phase transition and the ratio of the bulk
viscosity to the entropy density $\xi/s$ can have a maximum near
the phase transition for some values of the vacuum sigma mass if
the peak in the $\xi$ is sharp enough.

The structure of the paper is the following. Sec. \ref{hardcorsec}
contains the constant cross sections model description. The
applicability of the used approximations is discussed in Sec.
\ref{condappl}. The Boltzmann equation, its solution and general
form of the transport coefficients can be found in Sec.
\ref{CalcSec}. The numerical calculations for the hadron gas are
presented in Sec. \ref{numcalc}. In Sec. \ref{singcomsec}
analytical results for the single-component gas are presented. In
particular analytical expression for the first order
single-component shear viscosity coefficient, found before in
\cite{anderson}, is corrected while the bulk viscosity coefficient
remains the same. Also the nonequilibrium distribution function in
the same approximation is written. Some analytical results for the
binary mixture are considered in Sec. \ref{binmixsec}. Integrals
of source terms needed for calculation of the transport
coefficients can be found in Appendix \ref{appA}. The general
entropy density formula used in numerical calculations for the
hadron gas can be found in Appendix \ref{appTherm}.
Transformations of the collision brackets being the 12-dimensional
integrals which enter in transport coefficients can be found in
Appendix \ref{appJ}. The closed-form expressions for collision
rates, mean free paths and mean free times are included in
Appendix \ref{appmfp}.

\section{ The hard core interaction model \label{hardcorsec}}

In non-relativistic classical theory of particle interactions
there is a widespread model called the hard core repulsion model
or model of hard spheres with some radius $r$. For applications
for the high-energy nuclear collisions see
\cite{Gorenstein:2007mw} and references therein. The differential
scattering cross section for this model can be inferred from the
problem of scattering of point particle on the spherical potential
${U(r)=\infty}$ if ${r\leq a}$ and ${U(r)=0}$ if ${r>a}$
\cite{landau1}. In this model the differential cross section is
equal to $a^2/4$. To apply this result to the gas of hard spheres
with radius $r$ one can notice that the scattering of two spheres
can be considered as the scattering of the point particle on the
sphere of the radius $2r$, so that one should take ${a=2r}$. The
full cross section is obtained after integration over angles of
the $r^2 d\Omega$ which results in the $4\pi r^2$. For collisions
of hard spheres of different radiuses one should take
${a=r_1+r_2}$ or replace $r$ on $\frac{r_1+r_2}2$. The
relativistic generalization of this model is the constant (not
dependent on the scattering energy and angle) differential cross
sections model.

The hard spheres model is classical and connection of its cross
sections to cross sections calculated in any quantum theory is
needed. For particles having spin the differential cross sections
averaged over the initial spin states and summed over the final
ones will be used.\footnote{It's assumed that particle numbers of
the same species but with different spin states are equal. If this
were not so then in approximation, in which spin interactions are
neglected and probabilities to have certain spin states are equal,
the numbers of the particles with different spin states would be
approximately equal in the mean free time. With equal particle
numbers theirs distribution functions are equal too. This allows
one to use summed over the final states cross sections in the
Boltzmann equation.} If colliding particles are identical and
their differential cross section is integrated over the momentums
(or the spatial angle to get the total cross section) then it
should be multiplied on the factor $\frac12$ to cancel double
counting of the momentum states. These factors are exactly the
factors $\gamma_{kl}$ next to the collision integrals in the
Boltzmann equations (\ref{boleqs}). The differential cross
sections times these factors will be called the classical
differential cross sections.

From the analysis of semi-empirical total elastic cross sections
in \cite{prakash} for $\pi\pi$, $\pi K$, $\pi N$, $KN$ and $NN$
collisions one can find that the elastic part without resonance
contribution of the total cross sections lies approximately in the
range ${10-30~mb=1-3~fm^2}$ (except for $NN$ cross section
reaching $50~mb$ for small energies). From comparison of these
values with the total cross sections in the hard spheres model one
finds for the radius ${r=0.28-0.49~fm}$. For simplification the
mean value is chosen equal to ${r=0.4~fm}$ in numerical
calculations for all hadron and resonance elastic cross sections.
Then classical differential cross sections become equal to some
one constant value too. In this approximation the cross section
will enter in the transport coefficients as one factor.

\section{ Conditions of applicability \label{condappl}}

First the applicability of the Boltzmann equation and of
calculations of transport coefficients should be clarified.

Although the Boltzmann equation is valid for any perturbations of
the distribution function it should be a slowly varying function
of space-time coordinate to justify that it can be considered as a
function of macroscopic quantities like temperature, chemical
potential or flow velocity or in other words that one can apply
thermodynamics locally. Then one can make expansion over
independent gradients of thermodynamic functions and flow velocity
(the Chapman-Enskog method), which vanish in equilibrium.
Smallness of these perturbations of the distribution functions in
compare to theirs main parts ensures the validity of this
expansion and that the gradients are small.\footnote{The
magnitudes of thermodynamic quantities can also be restricted by
this condition or, conversely, not restricted even if transport
coefficients diverge. See also Sec. \ref{singcomsec} of this
paper. The smallness of the shear and the bulk viscosity gradients
can also be checked by the condition of smallness of the
$T^{(1)\mu\nu}$ (\ref{T1}) in compare to the $T^{(0)\mu\nu}$
(\ref{T0}).} Because these perturbations are inversely
proportional to coupling constants one can say that they are
proportional to some product of particles' mean free paths and the
gradients. So that in other words the mean free paths should be
much smaller than the characteristic lengths on which macroscopic
quantities change considerably\footnote{ It's clear that the mean
free paths should be smaller than the system's size too. }.

The system with number-changing processes should be treated
additionally. Number-changing processes are very important for the
bulk viscosity. If the coupling constant of inelastic
number-changing processes is small then the bulk viscosity is
dominated by the inelastic processes \cite{jeon}. Formally tending
coupling constants of inelastic processes to zero the bulk
viscosity diverges together with the nonequilibrium perturbations
of the distribution functions, which should be small. Though this
dependence can become invalid earlier if inelastic processes are
negligible in some sense, because the limiting case with
completely switched off inelastic processes can be well defined.
That's why there is a need to specify reasonable conditions when
inelastic processes can be neglected. Some simple ones of them are
proposed below. It is smallness of the time $t$ on which the
system's evolution is considered in compare to the mean free time
for number-changing processes of the $k'$-th\footnote{Primed
indexes run over particle species without regard to their spin
states. This assignment is clarified more below.} particle
species, $t^{inel}_{k'}$ (\ref{tkinel}),
 \eq{
 \frac{t}{t^{inel}_{k'}}=t R^{inel}_{k'}\ll 1,
 }
where $R^{inel}_{k'}$ is the rate of inelastic processes per
particle of the $k'$-th species. Similar condition, stating that
the number of reactions is smaller than unity, can be used:
 \eq{
 t V \sum_{n\in~\text{all channels}}\widetilde R^{inel}_{k',n}< 1,
 }
where $V$ is the system's volume and $\sum_{n\in~\text{all
channels}}\widetilde R^{inel}_{k',n}$ is the number of the
inelastic reactions of particles of the $k'$-th species over all
channels per unit time per unit volume. Though in mixtures with
mean free times of inelastic processes close to each other one
might need to use relevant sums of reaction rates instead of
reaction rates for certain species. Also one can consider natural
time-scales like relaxation times of the gradients or
thermodynamic functions. Estimations of the thermal and chemical
relaxation times for pions were done in \cite{Song:1996ik}. Basing
on these results one can expect that the approximate temperature
where inelastic processes cease to be efficient is
${T=120-140~MeV}$ for vanishing chemical potentials. For nonzero
chemical potentials this temperature is expected to be smaller.

Errors due to the Boltzmann statistics used instead of the Bose or
the Fermi ones were found to be small for vanishing chemical
potentials.\footnote{It should be mentioned that if particles of
the $k$-th particle species are bosons and if $\mu_k(x^\mu) \geq
m_k$ then there is (local) Bose-Einstein condensation for them
which should be treated in a special way.} According to
calculations for the pion gas in \cite{davesne} the bulk viscosity
becomes $25\%$ larger at ${T=120~MeV}$ and $33\%$ larger at
${T=200~MeV}$ for vanishing chemical potential. Although the
relative deviations of the thermodynamic quantities of the pion
gas at nonvanishing chemical potential ${\mu=100~MeV}$ are not
more than $20\%$\footnote{The relative deviations of the
thermodynamic quantities grow with the temperature for some fixed
value of the chemical potential and tend to some constant.} the
bulk viscosity becomes up to $2.5$ times more. The shear viscosity
becomes $15\%$ less at ${T=120~MeV}$ and $25\%$ less at
${T=200~MeV}$ for vanishing chemical potential and $33\%$ less at
${T=120~MeV}$ and $67\%$ less at ${T=200~MeV}$ for the
${\mu=100~MeV}$. The corrections to the bulk viscosity of the
fermion gas, according to calculations of the bulk viscosity
source term not presented in this paper, are of the opposite sign
and approximately of the same magnitude.

The condition of applicability of the ideal gas equation of state
is controlled by the dimensionless parameter $\upsilon n$ which
appears in the first correction from the binary collisions in the
virial expansion and should be small. Here ${\upsilon=16\pi
r^3/3}$ is the so called excluded volume parameter and $1/n$ is
the mean volume per particle. One finds ${\upsilon n\approx 0.09}$
at ${T=120~MeV}$ and ${\upsilon n\approx 0.2}$ at ${T=140~MeV}$
for vanishing chemical potentials. Though even small corrections
to thermodynamic quantities can result in large corrections for
the bulk viscosity as it was found for the quantum statistics
corrections. Along the freeze-out line (its parameters can be
found in \cite{Gorenstein:2007mw}) the $\upsilon n$ grows from
$0.07$ to $0.49$ with the temperature.

One more important requirement which one needs to justify the
Boltzmann equation approach is that the mean free time should be
much larger than $\hbar/\Omega$ ($\Omega$ is the characteristic
single-particle energy) \cite{danielewicz} or de Broglie
wavelength should be much smaller than the mean free path
\cite{Arnold:2002zm} to distinguish independent acts of collisions
and for particles to have well-defined on-shell energy and
momentum. This condition gets badly satisfied for high
temperatures or densities. The mean free path of particle species
$k'$ is given by the formula (\ref{mfp}) or the formula
(\ref{lel}) if inelastic processes can be neglected. The
wavelength can be written as ${\lambda_{k'}=\frac1{\langle
\abs{\vec p_{k'}}\rangle}}$, where the averaged modulus of
momentum of $k'$-th species $\langle \abs{\vec p_{k'}}\rangle$ is
 \eq{\label{avmom}
 \langle\abs{\vec p_{k'}}\rangle=\frac{\int d^3p_{k'} |\vec p_{k'}|
 f^{(0)}_{k'}(p_{k'})}{\int d^3p_{k'} f^{(0)}_{k'}(p_{k'})}=\frac{2
 e^{-z_{k'}} T (3 + 3z_{k'} + z_{k'}^2)}{z_{k'}^2
 K_2(z_{k'})}=\sqrt{\frac{8 m_{k'} T}{\pi}}
 \frac{K_{5/2}(z_{k'})}{K_2(z_{k'})},
 }
where ${z_{k'}\equiv m_{k'}/T}$, $K_2(x)$ is the modified Bessel
function of the second kind. As it follows from the (\ref{avmom})
the largest wavelength is for the lightest particles,
$\pi$-mesons. The elastic collision mean free paths are close to
each other for all particle species. Hence, the smallest value of
the ratio $\lambda_{k'}/l^{el}_{k'}$ is for $\pi$-mesons. Its
value is close to the value of the $\upsilon n$ and is
exponentially suppressed for small temperatures too. At
temperature ${T=140~MeV}$ and vanishing chemical potentials this
ratio is equal to $0.18$. Along the freeze-out line it grows from
$0.12$ to $0.37$ with the temperature.

If the last requirement is not satisfied one can use the Kubo
formulas, for instance. In \cite{jeon} it was shown that
intermediate integral equations for the shear and the bulk
viscosities coming after linearization of the Boltzmann equation
and expansion over gradients of the flow velocity can be
reproduced from the Kubo formulas in scalar
${g\phi^3+\lambda\phi^4}$ theory. The difference is in that one
has to replace particles' masses with temperature dependent ones
and to use thermal matrix elements for elastic and inelastic
collisions. Basing on this result an example of effective kinetic
theory of quasiparticle excitations valid for all temperatures was
presented in \cite{jeonyaffe}. For further developments see
\cite{Arnold:2002zm}, \cite{Gagnon:2006hi}, \cite{Gagnon:2007qt}.
For other approaches see \cite{Blaizot:1992gn},
\cite{Calzetta:1986cq}, \cite{Calzetta:1999ps} and
\cite{Arnold:1997gh} with references therein.

\section{ Details of calculations \label{CalcSec} }
\subsection{ The Boltzmann equation and its solution }

Calculations in this paper go close to the ones in \cite{groot}
though with some differences and generalizations. Let's start from
some definitions. Multi-indices $k,l,m,n$ will be used to denote
particle species with certain spin states. Indexes $k',l',m',n'$
will be used to denote particle species without regard to their
spin states (and run from 1 to the number of particle species
$N'$) and $a,b$ to denote conserved quantum numbers\footnote{In
systems with only elastic collisions each particle species have
their own "conserved quantum number", equal to 1.}. Because
nothing depends on spin variables one has for every sum over the
multi-indexes
 \eq{
  \sum_k ... = \sum_{k'}g_{k'}...,
 }
where $g_{k'}$ is the spin degeneracy factor. The following
assignments will be used
 \begin{eqnarray}\label{assign1}
  \nonumber n\equiv\sum_{k}n_k&\equiv&\sum_{k'} n_{k'}, \quad n_a\equiv \sum_k q_{ak} n_k,
  \quad x_k\equiv\frac{n_k}{n}, \quad x_{k'}\equiv\frac{n_{k'}}{n}, \quad x_a\equiv\frac{n_a}{n}, \\
  \hat \mu_k&=&\frac{\mu_k}{T}, \quad \hat \mu_a=\frac{\mu_a}{T},
  \quad z_k\equiv\frac{m_k}{T}, \quad \pi_k^\mu\equiv \frac{p_k^\mu}{T},
  \quad \tau_k\equiv \frac{p_k^\mu U_\mu}{T},
 \end{eqnarray}
where $q_{ak}$ denotes values of conserved quantum numbers of the
$a$-th kind of the $k$-th particle species. Everywhere the
particle number densities are summed the spin degeneracy factor
$g_{k'}$ appears and then gets absorbed into the $n_{k'}$ or the
$x_{k'}$ by definition. All other quantities with primed and
unprimed indexes don't differ, except for rates, mean free times
and mean free paths defined in Appendix \ref{appmfp}, the
$\gamma_{kl}$ commented below, the coefficients $A_{k'l'}^{rs}$,
$C_{k'l'}^{rs}$ and, of course, quantities whose free indexes set
indexes of the particle number densities $n_k$. Also the
assignment ${\int \frac{d^3p_k}{p_k^0}\equiv \int_{p_k}}$ will be
used somewhere for compactness.

The particle number flows are\footnote{The $+,-,-,-$ metric
signature is used throughout the paper.}
 \eq{\label{pflow}
 N^{\mu}_k=\int \frac{d^3p_k}{(2\pi)^3p^0_k} p^\mu_k f_k,
 }
where the assignment ${f_k(p_k)\equiv f_k}$ is introduced. The
energy-momentum tensor is
 \eq{\label{enmomten}
 T^{\mu\nu}=\sum_k \int \frac{d^3p_k}{(2\pi)^3p_k^0}p_k^\mu p_k^\nu f_k.
 }
The local equilibrium distribution functions are
 \eq{\label{loceq}
 f^{(0)}_k=e^{(\mu_k-p_k^\mu U_\mu)/T},
 }
where $\mu_k$ is the chemical potential of the $k$-th particle
species, $T$ is the temperature and $U_\mu$ is the relativistic
flow 4-velocity such that ${U_\mu U^\mu=1}$ (with frequently used
consequence ${U_\mu\p_\nu U^\mu=0}$). The local equilibrium is
considered as perturbations of independent thermodynamic variables
and the flow velocity over a global equilibrium such that they can
depend on the space-time coordinate $x^\mu$. Though such
perturbations are not the general ones and do not take into
account all possible deviations from a chemical equilibrium. For
numerical calculations along the freeze-out line, where such
deviations are important, saturation $\gamma$-factors are used,
see \cite{Cleymans:2005xv}, \cite{Andronic:2005yp} and references
therein. The chemical equilibrium implies that the particle number
densities are equal to their global equilibrium values. The global
equilibrium is called the time-independent stationary state with
the maximal entropy\footnote{There is also a kinetic equilibrium
which implies that the momentum distributions are the same as in
the global equilibrium. Thus, a state of a system with both the
pointwise (for the whole system) kinetic and the pointwise
chemical equilibria is the global equilibrium.}. The global
equilibrium of isolated system can be found by variation of the
total nonequilibrium entropy functional \cite{landau5} over the
distribution function with condition of the total energy and the
total net charges conservation:
 \eq{
 U[f]=\sum_k \int \frac{d^3p_kd^3x}{(2\pi)^3}
 f_k(1-\ln f_k)-\sum_k\int \frac{d^3p_kd^3x}{(2\pi)^3}\beta
 p^0_kf_k-\sum_{a,k} \lambda_a q_{ak} \int \frac{d^3p_kd^3x}{(2\pi)^3} f_k,
 }
where $\beta, \lambda_a$ are Lagrange coefficients. Equating the
first variation to zero one easily gets the function (\ref{loceq})
with ${U^\mu=(1,0,0,0)}$, ${\beta=\frac1{T}}$ and
 \eq{
 \mu_k=\sum_a q_{ak}\mu_a,
 }
where ${\mu_a=\lambda_a}$ are the independent chemical potentials
coupled to conserved net charges.

With ${f_k=f_k^{(0)}}$ substituted in the (\ref{pflow}) and the
(\ref{enmomten}) one gets the leading contribution in the
gradients expansion of the particle number flow and the
energy-momentum tensor
 \eq{
 N^{(0)\mu}_k=n_kU^\mu,
 }
 \eq{\label{T0}
 T^{(0)\mu\nu}=\epsilon U^\mu U^\nu - P\Delta^{\mu\nu},
 }
where the projector
 \eq{\label{proj}
 \Delta^{\mu\nu}\equiv g^{\mu\nu}-U^\mu U^\nu,
 }
is introduced. The $n_k$ is the ideal gas particle number density
 \eq{\label{ignk}
 n_k=U_\mu N^{(0)\mu}_k=\frac{1}{2\pi^2}T^3z_k^2K_2(z_k)e^{\hat \mu_k},
 }
the $\epsilon$ is the ideal gas energy density
 \eq{\label{epsandek}
 \epsilon=U_\mu U_\nu T^{(0)\mu\nu}=\sum_k
 \int \frac{d^3p_k}{(2\pi)^3}p_k^0f_k^{(0)}=\sum_k n_k e_k, \quad
 e_k\equiv m_k\frac{K_3(z_k)}{K_2(z_k)}-T,
 }
and the $P$ is the ideal gas pressure
 \eq{
 P=-\frac13T^{(0)\mu\nu}\Delta_{\mu\nu}=\sum_k \frac13\int
 \frac{d^3p_k}{(2\pi)^3p_k^0}\vec{p_k}^2 f_k^{(0)}=\sum_k n_k T=nT.
 }
Also the following assignments are used
 \begin{eqnarray}\label{assign2}
  \nonumber e&\equiv&\frac{\epsilon}{n}=\sum_kx_ke_k, \quad h_k\equiv e_k+T,
  \quad h\equiv\frac{\epsilon+P}{n}=\sum_k x_kh_k, \\
  \hat e_k&\equiv&\frac{e_k}{T}=z_k\frac{K_3(z_k)}{K_2(z_k)}-1,
  \quad \hat e\equiv\frac{e}{T}, \quad \hat h_k\equiv\frac{h_k}{T}=
  z_k\frac{K_3(z_k)}{K_2(z_k)}, \quad \hat h\equiv \frac{h}{T}.
 \end{eqnarray}
Above $h$ is the enthalpy per particle, $e$ is the energy per
particle and $h_k$, $e_k$ are the enthalpy and the energy per
particle of the $k$-th particle species correspondingly which are
well defined in the ideal gas.

In relativistic hydrodynamics the flow velocity $U^\mu$ needs some
extended definition in relation to the thermodynamic quantities.
The most convenient condition applied to $U^\mu$ is the
Landau-Lifshitz condition \cite{landau6}. This condition states
that in the local rest frame (where the flow velocity is zero
though its gradient can have nonzero value) each fluid cell should
have zero momentum and its energy and net charge densities should
be related to other thermodynamic quantities through the
equilibrium thermodynamic relations (without contribution of
nonequilibrium dissipations). Its covariant mathematical
formulation is
 \eq{\label{lLcond}
 (T^{\mu\nu}-T^{(0)\mu\nu})U_\mu=0, \quad (N^\mu_a-N^{(0)\mu}_a)U_\mu=0.
 }
The next to leading correction over the gradients expansion to the
$T^{\mu\nu}$ can be written as expansion over the 1-st order
Lorentz covariant gradients which are rotationally and space
inversion invariant and satisfy the Landau-Lifshitz
condition\footnote{ Also this form of $T^{(1)\mu\nu}$ respects the
second law of thermodynamics \cite{landau6}. Implementation of the
Eckart condition would result in different form of the
$T^{(1)\mu\nu}$ \cite{groot}. } (\ref{lLcond}):
 \eq{\label{T1}
 T^{(1)\mu\nu}\equiv2\eta \overset{\circ}{\overline{\nabla^\mu
 U^\nu}}+\xi \Delta^{\mu\nu} \nabla_\rho U^\rho=\eta\left(\Delta^\mu_\rho \Delta^\nu_\tau
 +\Delta^\nu_\rho \Delta^\mu_\tau-\frac23\Delta^{\mu\nu}\Delta_{\rho\tau}\right)\nabla^\rho
 U^\tau+\xi \Delta^{\mu\nu} \nabla_\rho U^\rho,
 }
where for any tensor $a_{\mu\nu}$ the symmetrized traceless tensor
assignment is introduced
 \eq{\label{tracelessten}
 \overset{\circ}{\overline{a_{\mu\nu}}}\equiv \left(\frac{\Delta_{\mu\rho}
 \Delta_{\nu\tau}+\Delta_{\nu\rho} \Delta_{\mu\tau}}2-\frac13\Delta_{\mu\nu}
 \Delta_{\rho\tau}\right)a^{\rho\tau}\equiv \Delta_{\mu\nu\rho\tau}a^{\rho\tau},
 \quad \Delta^{\mu\nu}_{~~\rho\tau}\Delta^{\rho\tau}_{~~\sigma\lambda}=
 \Delta^{\mu\nu}_{~~\sigma\lambda}.
 }
The equation (\ref{T1}) is the definition of the shear $\eta$ and
the bulk $\xi$ viscosity coefficients. The $\xi \Delta^{\mu\nu}
\nabla_\rho U^\rho$ term in the (\ref{T1}) can be considered as
nonequilibrium contribution to the pressure which enters in the
(\ref{T0}).

By means of the projector (\ref{proj}) one can split the
space-time derivative $\p_\mu$ as
 \eq{
 \p_\mu=U_\mu U^\nu \p_\nu + \Delta_\mu^\nu\p_\nu \equiv U_\mu
 D+\nabla_\mu.
 }
where $D \equiv U^\nu \p_\nu$, $\nabla_\mu \equiv
\Delta_\mu^\nu\p_\nu$. In the local rest frame (where
${U^\mu=(1,0,0,0)}$) the $D$ becomes the time derivative and the
$\nabla_\mu$ becomes the spacial derivative. Then the Boltzmann
equations can be written in the form
 \eq{\label{boleqs}
 p_k^\mu\p_\mu f_k=(p_k^\mu U_\mu D + p_k^\mu \nabla_\mu )f_k
 =C_k^{el}[f_k]+C_k^{inel}[f_k],
 }
where $C_k^{inel}[f_k]$ represents the inelastic or
number-changing collision integrals (it is dropped in calculations
in this paper if the opposite is not stated explicitly) and
$C_k^{el}[f_k]$ is the elastic ${2\leftrightarrow2}$ collision
integral. The collision integral $C_k^{el}[f_k]$ has the form of
the sum of positive gain terms and negative loss terms. Its
explicit form is \footnote{The factor $\gamma_{kl}$ cancels double
counting in integration over momentums of identical particles. The
factor $\frac12$ comes from the relativistic normalization of
scattering amplitudes. } (cf. \cite{jeon, Arnold:2002zm})
 \begin{eqnarray}\label{ckel}
  \nonumber C_k^{el}[f_k]&=&\sum_{l} \gamma_{kl}\frac12\int
  \frac{d^3p_{1l}}{(2\pi)^32p_{1l}^0}\frac{d^3p'_k}{(2\pi)^32{p'}_k^0}
  \frac{d^3p'_{1l}}{(2\pi)^32{p'}_{1l}^0}(f'_{k}f'_{1l}-f_{k}f_{1l})\\
  &\times& |\mc M_{kl}|^2(2\pi)^4\delta^{4}(p'_k+p'_{1l}-p_k-p_{1l})
 \end{eqnarray}
where ${\gamma_{kl}=\frac12}$ if $k$ and $l$ denote the same
particle species without regard to spin states and
${\gamma_{kl}=1}$ otherwise, ${|\mc M_{kl} (p'_k,p'_{1l};
p_k,p_{1l})|^2 \equiv |\mc M_{kl}|^2}$ is the square of
dimensionless elastic scattering amplitude, averaged over the
initial spin states and summed over the final ones. Index $1$
designates that $p_k$ and $p_{1k}$ are different variables.
Introducing ${W_{kl}\equiv W_{kl}(p'_k,p'_{1l};p_k,p_{1l})}$ as
 \eq{
 W_{kl}=\frac{|\mc M_{kl}|^2}{64\pi^2}\delta^{4}(p'_k+p'_{1l}-p_k-p_{1l}),
 }
one can rewrite the collision integral (\ref{ckel}) in the form as
in \cite{groot}
 \eq{\label{ckelgroot}
 C_k^{el}[f_k]=(2\pi)^3\sum_{l}\gamma_{kl}
 \int_{p_{1l},{p'}_k,{p'}_{1l}}\left(\frac{f'_{k}}{(2\pi)^3}\frac{f'_{1l}}{(2\pi)^3}
 -\frac{f_{k}}{(2\pi)^3}\frac{f_{1l}}{(2\pi)^3}\right)W_{kl}.
 }
The $W_{kl}$ is related to the elastic differential cross section
$\sigma_{kl}$ as \cite{groot}
 \eq{
 W_{kl}=s\sigma_{kl}\delta^{4}(p'_k+p'_{1l}-p_k-p_{1l}),
 }
where ${s=(p_k+p_{1l})^2}$ is the usual Mandelstam variable. The
$W_{kl}$ has properties $W_{kl}(p'_k,p'_{1l};p_k,p_{1l}) =
W_{kl}(p_k,p_{1l};p'_{k},p'_{1l}) =
W_{lk}(p'_{1l},p'_{k};p_{1l},p_k)$ (due to time reversibility and
a freedom of relabelling of order numbers of particles taking part
in reaction). And e.g. $W_{kl}(p'_k,p'_{1l};p_k,p_{1l}) \neq
W_{kl}(p'_{1l},p'_{k};p_{1l},p_{k})$ in general case. Elastic
collision integrals have important properties which one can easily
prove \cite{groot}:
 \eq{\label{c22prop0}
 \int \frac{d^3p_k}{(2\pi)^3p^0_k} C_k^{el}[f_k]=0,
 }
 \eq{\label{c22prop}
 \sum_k\int \frac{d^3p_k}{(2\pi)^3p^0_k} p_k C_k^{el}[f_k]=0.
 }
Also the $C^{el}_k[f_k]$ vanishes if ${f_k=f^{(0)}_k}$.

The distribution functions $f_k$ solving the system of the
Boltzmann equations approximately are sought in the form
 \eq{\label{fpert}
 f_k=f^{(0)}_k+f^{(1)}_k\equiv f^{(0)}_k+f^{(0)}_k\varphi_k(x,p_k),
 }
where it's assumed that $f_k$ depend on the $x^\mu$ entirely
through the $T$, $\mu_k$, $U^\mu$ or their space-time derivatives.
Also it is assumed that ${|\varphi_k|\ll 1}$. After substitution
of ${f_k=f^{(0)}_k}$ in the (\ref{boleqs}) the r.h.s. becomes zero
and the l.h.s. is zero only if the $T$, $\mu_k$ and $U^\mu$ don't
depend on the $x^\mu$ (provided they don't depend on the momentum
$p^\mu_k$). The 1-st order space-time derivatives of the $T$,
$\mu_k$, $U^\mu$ in the l.h.s. should be cancelled by the first
nonvanishing contribution in the r.h.s. This means that the
$\varphi_k$ should be proportional to the 1-st order space-time
derivatives of the $T$, $\mu_k$, $U^\mu$. The covariant time
derivatives $D$ can be expressed through the covariant spacial
derivatives by means of approximate hydrodynamical equations,
valid at the same order in the gradients expansion. Let's derive
them. Integrating the (\ref{boleqs}) over the
$\frac{d^3p_k}{(2\pi)^3p^0_k}$ with the ${f_k=f^{(0)}_k}$ in the
l.h.s. with inelastic collision integrals retained and using the
(\ref{c22prop0}) and the (\ref{pflow}) one would get (which can be
justified with explicit form of inelastic collision integrals)
 \eq{\label{conteq}
 \p_\mu N^{(0)\mu}_k=Dn_k+n_k\nabla_\mu U^\mu=I_k,
 }
where $I_k$ is the sum of inelastic collision integrals integrated
over momentum. It is responsible for nonconservation of the total
particle number of the $k$-th particle species and has property
${\sum_k q_{ak} I_k=0}$. If ${C^{inel}_k[f_k]=0}$ then ${I_k=0}$
which results in conservation of the total particle numbers of
each particle species. Multiplying the (\ref{conteq}) on the
$q_{ak}$ and summing over $k$ one gets the continuity equations
for the net charge flows:
 \eq{\label{conteq2}
 \p_\mu N^{(0)\mu}_a=Dn_a+n_a\nabla_\mu U^\mu=0.
 }
Then integrating the (\ref{boleqs}) over the
$p_k^\mu\frac{d^3p_k}{(2\pi)^3p^0_k}$ with the ${f_k=f^{(0)}_k}$
in the l.h.s. one gets
 \eq{\label{encons0}
 \p_\rho T^{(0)\rho\nu}=\p_\rho(\epsilon U^\rho U^\nu-P\Delta^{\rho\nu})=0.
 }
There is zero in the r.h.s. even if inelastic collision integrals
are retained because they respect energy conservation too. Note
that the Boltzmann equations (\ref{boleqs}) permit self-consistent
consideration only of the ideal gas energy-momentum tensor and net
charge flows. After convolution of the (\ref{encons0}) with the
$\Delta^\mu_\nu$ one gets the Euler's equation
 \eq{\label{eulereq}
 DU^\mu=\frac1{\epsilon+P}\nabla^\mu P=\frac1{hn}\nabla^\mu P.
 }
After convolution of the (\ref{encons0}) with the $U_\nu$ one gets
equation for the energy density
 \eq{\label{encons1}
 D\epsilon=-(\epsilon+P)\nabla_\mu U^\mu = - hn\nabla_\mu U^\mu.
 }

To proceed further one needs to expand the l.h.s. of the Boltzmann
equations (\ref{boleqs}) over the gradients of thermodynamic
variables and the flow velocity. Let's choose $\mu_a$ and $T$ as
independent thermodynamic variables. Then for the $Df^{(0)}_k$ one
can write expansion
 \eq{\label{Dfk}
 Df^{(0)}_k=\sum_a \frac{\p f^{(0)}_k}{\p\mu_a}D\mu_a+\frac{\p f^{(0)}_k}{\p
 T}DT+\frac{\p f^{(0)}_k}{\p U^\mu}DU^\mu.
 }
Writing the expansion for the $Dn_a$ and the $D\epsilon$ one gets
from the (\ref{conteq2}) and the (\ref{encons1}):
 \eq{\label{Dna}
 Dn_a=\sum_b \frac{\p n_a}{\p \mu_b}D\mu_b+\frac{\p n_a}{\p T}DT
 =-n_a\nabla_\mu U^\mu,
 }
 \eq{\label{Depsilon}
 D\epsilon=\frac{\p\epsilon}{\p T}DT+\sum_a\frac{\p \epsilon}{\p \mu_a}D\mu_a
 =-hn\nabla_\mu U^\mu.
 }
The solution to the system of equations (\ref{Dna}),
(\ref{Depsilon}) can easily be found:
 \eq{\label{Teqn}
 DT=-R T\nabla_\mu U^\mu,
 }
 \eq{\label{mueqn}
 D\mu_a=T\sum_b \tilde{A}^{-1}_{ab}(R B_b-x_b)\nabla_\mu U^\mu,
 }
where
 \eq{\label{Rdef}
 R\equiv\frac{\hat h-\sum_{a,b} E_a \tilde{A}^{-1}_{ab}x_b}{C_{\{\mu\}}-\sum_{a,b}
 E_a\tilde{A}^{-1}_{ab}B_b},
 }
and
 \eq{
 \frac{\p n_a}{\p \mu_b}\equiv \frac{n}{T}\tilde{A}_{ab},
 \quad \frac{\p n_a}{\p T}\equiv \frac{n}{T}B_a, \quad
 \frac{\p\epsilon}{\p T}\equiv n C_{\{\mu\}},\quad
 \frac{\p\epsilon}{\p\mu_a}\equiv n E_a.
 }
Above it is assumed that the matrix $\tilde{A}_{ab}$ is not
degenerate, which is the case usually. Otherwise uncertainties or
singularities from $D\mu_a$ enter in the bulk viscosity. Using the
ideal gas formulas (\ref{ignk}) and (\ref{epsandek}) one gets
 \begin{eqnarray}\label{ABCE}
  \nonumber \tilde A_{ab}&=&\sum_k q_{ak}q_{bk}x_k, \quad
  E_a=\sum_k q_{ak} x_k \hat e_k, \quad B_a=E_a-\sum_b \tilde A_{ab}\hat\mu_b,\\
  C_{\{\mu\}}&=&\sum_k x_k(3\hat h_k+z_k^2-\hat \mu_k\hat
  e_k)=\sum_kx_k(3\hat h_k+z_k^2)-\sum_aE_a\hat\mu_a.
 \end{eqnarray}
From the (\ref{ABCE}) one can see that the matrix $\tilde A_{ab}$
has positive diagonal elements and in the case of one kind of
charge it's always not degenerate. For the special case of
vanishing chemical potentials, $\mu_a=0$, the quantities $n_a$,
$x_a$, $B_a$, $E_a$ tend to zero because the contributions from
particles and anti-particles cancel each other and chargeless
particles don't contribute. Then from the (\ref{Teqn}) and the
(\ref{mueqn}) one finds
 \eq{\label{Teqnmu0}
 DT|_{\mu_a=0}=-\frac{h}{C_{\{\mu\}}}\nabla_\mu U^\mu,
 }
 \eq{
 D\mu_a|_{\mu_a=0}=0.
 }
This means that for vanishing chemical potentials one can simply
exclude them from the distribution functions (if one does not
study diffusion and thermal conductivity). In systems with only
elastic collisions each particle has its own charge so that one
takes ${q_{ak}=\delta_{ak}}$ and gets
 \begin{eqnarray}\label{elquant}
  \nonumber \tilde A_{kl}&=&\delta_{kl}x_k, \quad B_k=x_k(\hat e_k-\hat \mu_k),
  \quad E_k=\hat e_k x_k, \quad R=\frac1{c_\upsilon} \\
  C_{\{\mu\}}&-&\sum_{a,b}E_a \tilde A^{-1}_{ab} B_{b}=\sum_kx_k(-\hat h_k^2+5\hat
  h_k+z_k^2-1)\equiv\sum_kx_k c_{\upsilon,k}\equiv c_\upsilon.
 \end{eqnarray}
Then the equation for the $DT$ (\ref{Teqn}) remains the same with
a new $R$ from the (\ref{elquant}) and the equations (\ref{mueqn})
become:
 \eq{\label{mukeqn}
 D\mu_k=\left(\frac{T}{c_\upsilon}(\hat e_k-\hat \mu_k)-T\right)\nabla_\mu U^\mu.
 }
Note that in systems with only elastic collisions the $D\mu_k$
does not tend to zero for vanishing chemical potentials so that
the $\mu_k$ could not be omitted in the distribution functions in
this case. Because the heat conductivity and diffusion are not
considered in this paper their nonequilibrium gradients are taken
equal to zero, $\nabla_\nu P=\nabla_\nu T=\nabla_\nu\mu_a=0$.
Using the (\ref{Teqn}), (\ref{mueqn}) and (\ref{eulereq}) the
l.h.s. of the (\ref{boleqs}) can be transformed as
 \eq{\label{boleqnlhs}
 (p_k^\mu U_\mu D+p_k^\mu\nabla_\mu)f_k^{(0)} = -Tf_k^{(0)}
 \pi_k^\mu \pi_k^\nu \overset{\circ}{\overline{\nabla_\mu
 U_\nu}}+Tf_k^{(0)}\hat Q_k\nabla_\rho U^\rho,
 }
where
 \eq{\label{Qsource}
 \hat Q_k\equiv \tau_k^2\left(\frac13-R\right)+\tau_k\left[\sum_{a,b}q_{ak} \tilde
 A^{-1}_{ab}(RB_b-x_b)+\hat\mu_k R\right]-\frac13z_k^2.
 }
Using the (\ref{tracelessten}) one can notice that the useful
equality ${ \pi_k^\mu \pi_k^\nu \overset{\circ}{
\overline{\nabla_\mu U_\nu} } = \overset{\circ}{ \overline{
\pi_k^\mu \pi_k^\nu } }\overset{\circ}{ \overline{ \nabla_\mu
U_\nu } } }$ holds. In systems with only elastic collisions the
$\hat Q_k$ simplifies in agreement with \cite{groot}:
 \eq{\label{Qsource2}
 \hat Q_k=\left(\frac43-\gamma\right)\tau_k^2+
 \tau_k((\gamma-1)\hat h_k-\gamma)-\frac13z_k^2.
 }
where the assignments $\gamma$ from \cite{groot} is used. It can
be expressed through the $c_\upsilon$, defined in the
(\ref{elquant}), as ${\gamma\equiv \frac1{c_\upsilon}+1}$.
Introducing symmetric round brackets
 \eq{
 (F,G)_k\equiv\frac1{4\pi z_k^2K_2(z_k)T^2}\int_{p_k} F(p_k)G(p_k)e^{-\tau_k}.
 }
and assignments
 \eq{
 \alpha_k^r\equiv(\hat Q_k,(\tau_k)^r), \quad \gamma_k^r\equiv((\tau_k)^r
 \overset{\circ}{\overline{\pi_k^\mu\pi_k^\nu}},
 \overset{\circ}{\overline{\pi_{k\mu}\pi_{k\nu}}}), \quad
 a^r_k\equiv(1,(\tau_k)^r)_k,
 }
and using explicit expressions of the $a_k^r$ from Appendix
\ref{appA} one finds for the $\alpha_k^0$ and the $\alpha_k^1$ in
systems with elastic and inelastic collisions:
 \eq{\label{alphak0}
 \alpha_k^0=1+\sum_{a,b}q_{ak} \tilde A^{-1}_{ab}(R B_b-x_b)-(\hat e_k-\hat\mu_k)R,
 }
 \eq{\label{alphak1}
 \alpha_k^1=\hat h_k+\sum_{a,b}\hat e_k q_{ak}\tilde A^{-1}_{ab}(R
 B_b-x_b)-(3\hat h_k+z_k^2+\hat\mu_k(1-\hat h_k))R.
 }
Then using the (\ref{alphak0}) and the (\ref{alphak1}) one can
show that
 \eq{\label{lhsnchcons}
 \sum_kq_{ak}x_k\alpha_k^0=0,
 }
 \eq{\label{lhsencons}
 \sum_kx_k\alpha_k^1=0.
 }
Because the gradients $\nabla_\mu U^\mu$ and
$\overset{\circ}{\overline{\nabla_\mu U_\nu}}$ are independent the
(\ref{lhsnchcons}) and the (\ref{lhsencons}) are direct
consequences of the local net charge (\ref{conteq2}) and the
energy-momentum (\ref{encons0}) conservations. Quantities
$(1,\overset{\circ}{\overline{\pi^\mu_{k}\pi^\nu_{k}}})$ and
$(p_k^\lambda,\overset{\circ}{\overline{\pi^\mu_{k}\pi^\nu_{k}}})$
vanish automatically because of the special tensorial structure of
the $\overset{\circ}{ \overline{ \pi^\mu_{k} \pi^\nu_{k}}
}$.\footnote{ Direct computation gives $(1,\overset{\circ}{
\overline{\pi^\mu_k \pi^\nu_k} })_k \propto (C_1 U^\sigma U^\rho +
C_2\Delta^{\sigma\rho}) \Delta_{ ~~\sigma\rho }^{\mu\nu} = 0$,
$(p_k^\lambda, \overset{\circ}{ \overline{\pi^\mu_k \pi^\nu_k}
})_k \propto (C_1 U^\lambda U^\sigma U^\rho + C_2U^\lambda
\Delta^{\sigma\rho} + C_3U^\sigma \Delta^{\lambda\rho})
\Delta_{~~\sigma\rho}^{\mu\nu} = 0$.}

The next step is to transform the r.h.s of the Boltzmann equations
(\ref{boleqs}). After substitution of the (\ref{fpert}) in the
r.h.s. of the (\ref{boleqs}) the collision integral becomes linear
and one gets
 \eq{\label{boleqnrhs}
 C_k^{el}[f_k]\approx -f_k^{(0)}\sum_l \mc L_{kl}^{el}[\varphi_k],
 }
where
 \eq{
 \mc L_{kl}^{el}[\varphi_k]\equiv\frac{\gamma_{kl}}{(2\pi)^3}\int_{p_{1l},{p'}_k,{p'}_{1l}}
 f_{1l}^{(0)}(\varphi_k+\varphi_{1l}-\varphi'_k-\varphi'_{1l})W_{kl},
 }
The unknown function $\varphi_k$ is sought in the form
 \eq{\label{varphi}
 \varphi_k=\frac1{n\sigma(T)}\left(-A_k(p_k)\nabla_\mu U^\mu+C_k(p_k)
 \overset{\circ}{\overline{\pi^\mu_k \pi^\nu_k}}
 \overset{\circ}{\overline{\nabla_\mu U_\nu}}\right),
 }
where $\sigma(T)$ is some formal averaged cross section, used to
come to dimensionless quantities. Then using the (\ref{boleqnlhs})
and the (\ref{boleqnrhs}) and the fact that the gradients
$\nabla_\mu U^\mu$ and $\overset{\circ}{\overline{\nabla_\mu
U_\nu}}$ are independent the Boltzmann equations can be written as
independent integral equations:
 \eq{\label{xieqn}
 \hat Q_k=\sum_l x_l L_{kl}^{el}[A_k],
 }
 \eq{\label{etaeqn}
 \overset{\circ}{\overline{\pi_{k}^\mu \pi_{k}^\nu}}=
 \sum_l x_l L_{kl}^{el}[C_k \overset{\circ}{\overline{\pi_{k}^\mu \pi_{k}^\nu}}],
 }
where the dimensionless collision integral is introduced
 \eq{
 L_{kl}^{el}[\chi_k]=\frac1{n_lT\sigma(T)}\mc L_{kl}^{el}[\chi_k].
 }
In case of present inelastic processes the l.h.s of the
(\ref{xieqn}) is set by the source term (\ref{Qsource}) and the
r.h.s. contains linear inelastic collision integrals. After
introduction of inelastic processes the source term in the
(\ref{xieqn}) becomes much larger as demonstrated in Sec.
\ref{singcomsec}. Using the equations (\ref{mueqn}) and
(\ref{Teqn}) and the ideal gas formulas (\ref{ABCE}) one can check
that in the zero masses limit the $\hat Q_k$ (\ref{Qsource}) tend
to zero and ${D\hat\mu_a=0}$ that is the $\hat\mu_a$ don't scale
and the distribution functions become scale invariant. The source
term of the shear viscosity in the (\ref{etaeqn}) doesn't depend
on the presence of inelastic processes in the system and
originates from the free propagation term ${\frac{\vec p_k}{p_k^0}
\frac{\p f_k}{\p \vec r}}$ in the Boltzmann equation.

\subsection{ The transport coefficients and their properties }

After substitution of the $f_k^{(1)}$ with the $\varphi_k$
(\ref{varphi}) into (\ref{enmomten}) and comparison with the
(\ref{T1}) one finds the formula for the bulk viscosity
 \eq{\label{bulkvisc}
 \xi=-\frac13\frac{T}{\sigma(T)}\sum_k x_k(\Delta^{\mu\nu}\pi_{\mu,k}\pi_{\nu,k},A_k)_k,
 }
and for the shear viscosity
 \eq{\label{shearvisc}
 \eta=\frac1{10}\frac{T}{\sigma(T)}\sum_k x_k(\overset{\circ}{\overline{\pi^\mu_k \pi^\nu_k}},
 C_k \overset{\circ}{\overline{\pi_{k\mu} \pi_{k\nu}}})_k,
 }
where the relation ${\Delta^{\mu\nu}_{ ~~\sigma\tau }
\Delta_\mu^\sigma \Delta_\nu^\tau = 5}$ is used.

In kinetics the conditions, that the nonequilibrium perturbations
of distribution functions does not contribute to the net charge
and the energy-momentum densities, are used as convenient choice
and are called conditions of fit. They reproduce the
Landau-Lifshitz condition (\ref{lLcond}). The conditions of fit
for the net charge densities can be written as
 \eq{\label{cofchf}
 \sum_k q_{ak} \int \frac{d^3p_k}{(2\pi)^3p_k^0}p_k^\mu U_\mu
 f_k^{(0)}\varphi_k=0,
 }
and for the energy-momentum density can be written as
 \eq{\label{cofemt}
 \sum_k \int \frac{d^3p_k}{(2\pi)^3p_k^0}p_k^\mu p_k^\nu U_\nu f_k^{(0)}\varphi_k=0.
 }
For the special tensorial functions $C_k \overset{\circ}{
\overline{\pi_{k\mu} \pi_{k\nu}}}$ in the (\ref{varphi}) they are
satisfied automatically and for the scalar functions $A_k$ they
can be rewritten in the form (the 3-vector part of the
(\ref{cofemt}) is automatically satisfied)
 \eq{\label{condfit}
 \sum_k q_{ak}x_k (\tau_k, A_k)_k=0, \quad \sum_k x_k (\tau_k^2, A_k)_k=0.
 }
For a single-component gas with only elastic processes the
conditions (\ref{cofchf}) and (\ref{cofemt}) exclude zero modes
that is nonphysical solutions, proportional to the $p^\mu$ and a
constant, for which elastic collision integral vanishes. In a
single-component gas with inelastic collisions a constant is not a
zero mode. For a multi-component gas these conditions would just
modify the functional space on which solutions are sought. With
help of these conditions of fit one can show explicitly essential
positiveness of the $\xi$. Namely, using the conditions of fit
(\ref{condfit}), the equation (\ref{xieqn}) and the identity
${\Delta^{\mu\nu} \pi_{\mu,k}\pi_{\nu,k} = z_k^2 - \tau_k^2}$ the
bulk viscosity (\ref{bulkvisc}) can be rewritten as
 \eq{\label{xipos}
 \xi=\frac{T}{\sigma(T)}\sum_k x_k(\hat Q_k,A_k)_k=
 \frac{T}{\sigma(T)}\sum_k x_k\left(\sum_l x_l L^{el}_{kl}[A_k],A_k\right)_k=
 \frac{T}{\sigma(T)}[\{A\},\{A\}],
 }
where the square brackets are introduced for sets of equal lengths
${\{F\}=(F_1,...,F_k,...)}$, ${\{G\}=(G_1,...,G_k,...)}$:
 \eq{\label{sqrbra}
 [\{F\},\{G\}]\equiv\frac1{n^2\sigma(T)}\sum_{k,l}
 \frac{\gamma_{kl}}{(2\pi)^6}\int_{p_k,p_{1l},{p'}_k,{p'}_{1l}}
 f^{(0)}_kf^{(0)}_{1l}(F_k+F_{1l}-{F'}_{k}-{F'}_{1l})G_kW_{kl}.
 }
Using the time reversibility property of the $W_{kl}$ one can show
that equality
 \eq{
 (F_k+F_{1l}-{F'}_{k}-{F'}_{1l})G_k=\frac14(F_k+F_{1l}-{F'}_{k}-{F'}_{1l})
 (G_k+G_{1l}-{G'}_{k}-{G'}_{1l}),
 }
holds under integration and summation in the (\ref{sqrbra}). Then
one gets the direct consequence
 \eq{
 [\{F\},\{G\}]=[\{G\},\{F\}], \quad [\{F\},\{F\}]\geq 0.
 }
This proves the essential positiveness of the $\xi$. Similarly
using the (\ref{etaeqn}) the shear viscosity can be rewritten in
essentially positive form
 \begin{eqnarray}\label{etapos}
  \nonumber\eta&=&\frac1{10}\frac{T}{\sigma(T)}\sum_kx_k
  \left(\overset{\circ}{\overline{\pi_{k}^\mu \pi_{k}^\nu}},
  C_k \overset{\circ}{\overline{\pi_{k\mu} \pi_{k\nu}}}\right)_k
  =\frac1{10}\frac{T}{\sigma(T)}\sum_kx_k\left(\sum_l x_l L^{el}_{kl}
  [C_k \overset{\circ}{\overline{\pi_{k}^\mu \pi_{k}^\nu}}],
  C_k \overset{\circ}{\overline{\pi_{k\mu} \pi_{k\nu}}}\right)_k\\
  &=&\frac1{10}\frac{T}{\sigma(T)}[\{C \overset{\circ}{\overline{\pi^\mu \pi^\nu}}\},
  \{C \overset{\circ}{\overline{\pi_{\mu} \pi_{\nu}}}\}].
 \end{eqnarray}

The considered variational method allows to find approximate
solution of the integral equations (\ref{xieqn}) and
(\ref{etaeqn}) in the form of linear combination of test-functions
with coefficients, found from condition to deliver extremum to
some functional, which is equivalent to solving the integral
equations. One could take this functional in the form of some
special norm as in \cite{groot}. Or one can take somewhat
different functional like in \cite{Arnold:2003zc}, which is more
convenient, and get the same result. This generalized functional
can be written in the form
 \eq{\label{genfunctional}
 F[\chi]=\sum_k x_k(S_k^{\mu...\nu},\chi_{k\mu...\nu})_k-
 \frac12[\{\chi_k^{\mu...\nu}\},\{\chi_{k\mu...\nu}\}],
 }
where ${S_k^{\mu...\nu}=\hat Q_k}$ and ${\chi_{k\mu...\nu}=A_k}$
for the bulk viscosity and ${S_k^{\mu...\nu} = \overset{\circ}{
\overline{\pi_{k}^\mu \pi_{k}^\nu} }}$, ${\chi_k^{\mu...\nu} = C_k
\overset{\circ}{ \overline{\pi_{k}^\mu \pi_{k}^\nu} }}$ for the
shear viscosity. Equating to zero the first variance of the
(\ref{genfunctional}) over the $\chi_{k\mu...\nu}$ one gets
 \eq{
 \sum_kx_k(S_k^{\mu...\nu},\delta\chi_{k\mu...\nu})_k-
 [\{\chi_k^{\mu...\nu}\},\{\delta\chi_{k\mu...\nu}\}]=0.
 }
Because variations $\delta\chi_{k\mu...\nu}$ are arbitrary and
independent the generalized integral equation follows then:
 \eq{\label{genvareqn}
 S_k^{\mu...\nu}=\sum_l x_l L^{el}_{kl}[\chi_{k\mu...\nu}].
 }
The second variation of the (\ref{genfunctional}) is
 \eq{
 \delta^2 F[\chi]=-[\{\delta\chi_k^{\mu...\nu}\},\{\delta\chi_{k\mu...\nu}\}]\leq 0,
 }
which means that solution of the integral equations (\ref{xieqn})
and (\ref{etaeqn}) is reduced to variational problem of finding
the maximum of the functional (\ref{genfunctional}). Using the
(\ref{genvareqn}) the maximal value of the (\ref{genfunctional})
can be written as
 \eq{
 F_{max}[\chi]=\frac12[\{\chi_k^{\mu...\nu}\},\{\chi_{k\mu...\nu}\}]|_{\chi=\chi_{\max}}.
 }
Then using the (\ref{xipos}) and the (\ref{etapos}) one can write
the bulk and the shear viscosities through the maximal value of
the $F[\chi]$
 \eq{
 \xi=\left.2\frac{T}{\sigma(T)}F_{max}\right|_{S_k^{\mu...\nu}=\hat Q_k, \, \chi_{k\mu...\nu}=A_k},
 }
 \eq{
 \eta=\left.\frac15\frac{T}{\sigma(T)}F_{max}\right|_{S_k^{\mu...\nu}=
 \overset{\circ}{\overline{\pi_{k}^\mu \pi_{k}^\nu}}, \, \chi_k^{\mu...\nu}=
 C_k \overset{\circ}{\overline{\pi_{k}^\mu \pi_{k}^\nu}} }.
 }
This means that the precise solution of the (\ref{genvareqn})
delivers the maximal values for the transport coefficients.

The approximate solution of the system of the integral equations
(\ref{xieqn}) and (\ref{etaeqn}) are sought in the form
 \eq{\label{Atestf}
 A_{k}=\sum_{r=0}^{n_1} A_k^r\tau^r_k,
 }
 \eq{\label{Ctestf}
 C_{k}=\sum_{r=0}^{n_2} C_k^r\tau^r_k,
 }
where $n_1$ and $n_2$ set the number of used test-functions.
Test-functions used in \cite{Arnold:2003zc} would cause less
significant digit cancellation in numerical calculations but there
is a need to reduce the dimension of the 12-dimensional integrals
from these test-functions as more as possible to perform
calculations in reasonable time. The test-functions in the form of
just powers of momentums seem to be the most convenient for this
purpose. Questions concerning uniqueness and existence of solution
and convergence of the approximate solution to the precise one are
covered in \cite{groot}. As long as particles of the same particle
species but with different spin states are undistinguishable their
functions $\varphi_k$ (\ref{varphi}) are equal and the variational
problem is reduced to variation of coefficients $A_{k'}^r$ and
$C_{k'}^r$ and the bulk (\ref{xipos}) and the shear (\ref{etapos})
viscosities can be rewritten as
 \eq{\label{finxi}
 \xi=\frac{T}{\sigma(T)}\sum_{k'=1}^{N'}\sum_{r=0}^{n_1}x_{k'}\alpha_{k'}^r A_{k'}^r,
 }
 \eq{\label{fineta}
 \eta=\frac1{10}\frac{T}{\sigma(T)}\sum_{k'=1}^{N'}\sum_{r=0}^{n_2}x_{k'}\gamma_{k'}^rC_{k'}^r.
 }
After substitution of the approximate functions $A_{k'}$
(\ref{Atestf}) and $C_{k'}$ (\ref{Ctestf}) into the
(\ref{genfunctional}) and equating the first variation of the
functional to zero one gets the following matrix equations (with
multi-indexes ${(l',s)}$ and ${(k',r)}$) for the bulk and the
shear viscosities correspondingly\footnote{One can first derive
the same equations for the $A_k$ and $C_k$, treating them as
different functions for all $k$, with the coefficients
$A_{kl}^{rs}$ and $C_{kl}^{rs}$ having the same form as the
$A_{k'l'}^{rs}$ and $C_{k'l'}^{rs}$. Then after summation of
equations over spin states of identical particles and taking
${A_k=A_{k'}}$, ${C_k=C_{k'}}$ one reproduces the system of
equations for the $A_{k'}$ and $C_{k'}$.}
 \eq{\label{ximatreq}
 x_{k'}\alpha_{k'}^r=\sum_{l'=1}^{N'} \sum_{s=0}^{n_1} A_{l'k'}^{sr}A_{l'}^s,
 }
 \eq{\label{etamatreq}
 x_{k'}\gamma_{k'}^r=\sum_{l'=1}^{N'} \sum_{s=0}^{n_2} C_{l'k'}^{sr}C_{l'}^s,
 }
where introduced coefficients $A_{k'l'}^{rs}$ and $C_{k'l'}^{rs}$
are
 \eq{\label{A4ind}
 A_{k'l'}^{rs}=x_{k'}x_{l'}[\tau^r,\tau_1^s]_{k'l'}+\delta_{k'l'}x_{k'}\sum_{m'}
 x_{m'}[\tau^r,\tau^s]_{k'm'},
 }
 \eq{\label{C4ind}
 C_{k'l'}^{rs}=x_{k'}x_{l'}[\tau^r\overset{\circ}{\overline{\pi^{\mu} \pi^{\nu}}},
 \tau_1^s\overset{\circ}{\overline{\pi_{1\mu}\pi_{1\nu}}}]_{k'l'}+\delta_{k'l'}x_{k'}\sum_{m'}
 x_{m'}[\tau^r\overset{\circ}{\overline{\pi^{\mu} \pi^{\nu}}},
 \tau^s\overset{\circ}{\overline{\pi_{\mu} \pi_{\nu}}}]_{k'm'}.
 }
They are expressed through the collision brackets
 \eq{\label{br1}
 [F,G_1]_{kl}\equiv\frac{\gamma_{kl}}{T^6(4\pi)^2z_k^2z_l^2K_2(z_k)K_2(z_l)\sigma(T)}
 \int_{p_k,p_{1l},{p'}_k,{p'}_{1l}}e^{-\tau_k-\tau_{1l}}(F_k-{F'}_k)G_{1l}W_{kl}.
 }
The collision brackets $[F,G]_{kl}$ are obtained from the last
formula by replacement of the $G_{1l}$ on the $G_k$. Due to time
reversibility property of the $W_{kl}$ one can replace the
$G_{1l}$ on the ${\frac12(G_{1l}-{G'}_{1l})}$ in the (\ref{br1}).
Then one can see that
 \eq{\label{br2pos}
 [\tau^r,\tau^s]_{kl}>0.
 }
Also it's easy to notice the following symmetries
 \eq{
 [F,G_1]_{kl}=[G,F_1]_{lk}, \quad [F,G]_{kl}=[G,F]_{kl}.
 }
They result in the following symmetric properties ${A_{k'l'}^{rs}
= A_{l'k'}^{sr}}$, ${C_{k'l'}^{rs} = C_{l'k'}^{sr}}$. Also the
microscopical particle number and energy conservation laws imply
for the $A_{l'k'}^{sr}$:
 \eq{\label{A4chcons}
 A_{k'l'}^{0s}=0,
 }
 \eq{\label{A4encons}
 \sum_{k'}A_{k'l'}^{1s}=0.
 }
The (\ref{A4chcons}) together with the $\alpha_k^0=0$
(\ref{elal0}) means that the equations with ${r=0}$ in the
(\ref{ximatreq}) are excluded. From the (\ref{A4encons}) it
follows that each one equation with ${r=1}$ in the
(\ref{ximatreq}) can be expressed through the sum of the other
ones, reducing the rank of the matrix on 1. To solve the matrix
equation (\ref{ximatreq}) one eliminates one equation, for example
with ${k'=1}$, ${r=1}$. One of coefficients of $A_{l'}^1$ is
independent, for example, let it be $A_{1'}^1$. Using the
(\ref{A4encons}) the matrix equation (\ref{ximatreq}) can be
rewritten as
 \eq{\label{ximatreq2}
 x_{k'}\alpha_{k'}^r=\sum_{l'=2}^{N'}A_{l'k'}^{1r}(A_{l'}^1-A_{1'}^1)
 +\sum_{l'=1}^{N'} \sum_{s=2}^{n_1}A_{l'k'}^{sr}A_{l'}^s.
 }
Then using the (\ref{elal0}) and the (\ref{lhsencons}) the bulk
viscosity (\ref{finxi}) becomes
 \eq{\label{finxi2}
 \xi=\frac{T}{\sigma(T)}\sum_{k'=2}^{N'}x_{k'}\alpha_{k'}^1 (A_{k'}^1-A_{1'}^1)
 +\frac{T}{\sigma(T)}\sum_{k'=1}^{N'}\sum_{r=2}^{n_1}x_{k'}\alpha_{k'}^r A_{k'}^r.
 }
Then the coefficient $A_{1'}^1$ can be eliminated by shift of
other $A_{l'}^1$ and be implicitly used to satisfy one energy
conservation condition of fit. The particle number conservation
conditions of fit are implicitly satisfied by means of the
coefficients $A_{k'}^0$. The first term in the (\ref{finxi2}) is
present only in mixtures. That's why it is small in gases with
close to each other masses of particles of different species (like
in the pion gas). In gases with very different masses (like in the
hadron gas) contribution of the first term in the (\ref{finxi2})
can become dominant.

Analytical expressions for some lowest orders collision brackets
which enter in the matrix equations (\ref{etamatreq}) and
(\ref{ximatreq2}) can be found in Appendix \ref{appJ}. Higher
orders are not presented because of theirs bulky form.

\section{ The numerical calculations \label{numcalc}}
The numerical calculations for the hadron gas involve roughly
$2\frac{(N' n)^2}2$ 12-dimensional integrals, where $N'$ is the
number of particle species and $n$ is the number of used
test-functions (called the order of calculations). The
12-dimensional integrals being the collision brackets
${[F,G_1]_{kl}}$ and ${[F,G]_{kl}}$ can be reduced to
1-dimensional integrals, expressible through special functions. To
compute these special functions precisely and fast they were
expanded into series at several points. This allows to perform
calculations at the 3-rd and the 6-th orders for the shear and the
bulk viscosities correspondingly. Because analytical expressions
for the collision brackets are bulky the \emph{Mathematica}
\cite{math} was used for symbolical and some numerical
manipulations. The numerical calculations are done also for
temperatures above ${T=140~MeV}$, where inelastic collisions are
expected to play important role, for future comparisons and for
the bulk viscosity to show the position of its maximum in the
hadron gas.

\begin{figure}[h!]
\begin{center}
\epsfig{file=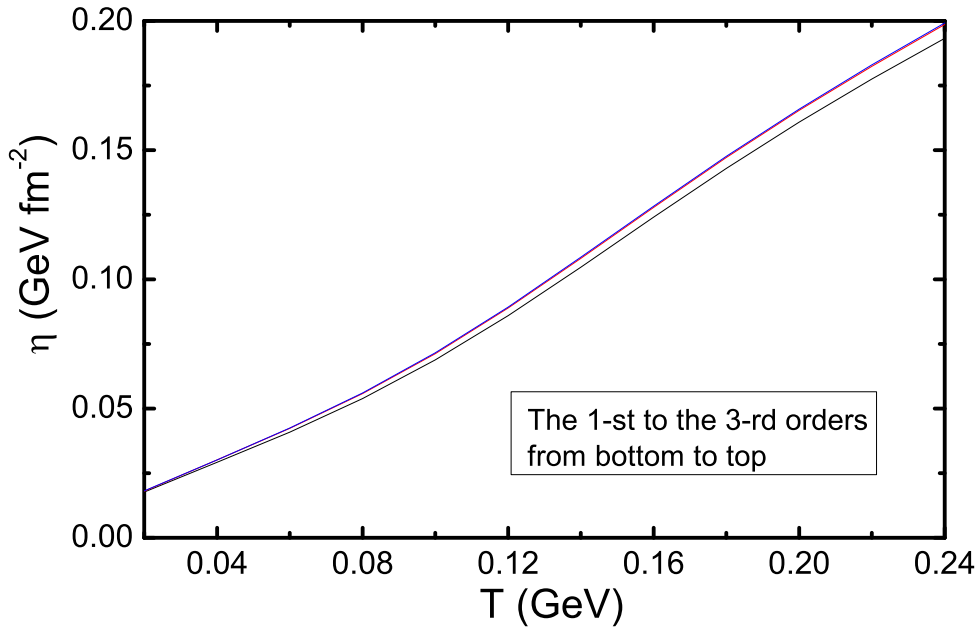,width=13cm} \caption{The shear
viscosity as function of the temperature in the hadron gas
calculated up to the 3-rd order. Chemical potentials are equal to
zero. \label{OrdersShearAll} }
\end{center}
\end{figure}

\begin{figure}[h!]
\begin{center}
\epsfig{file=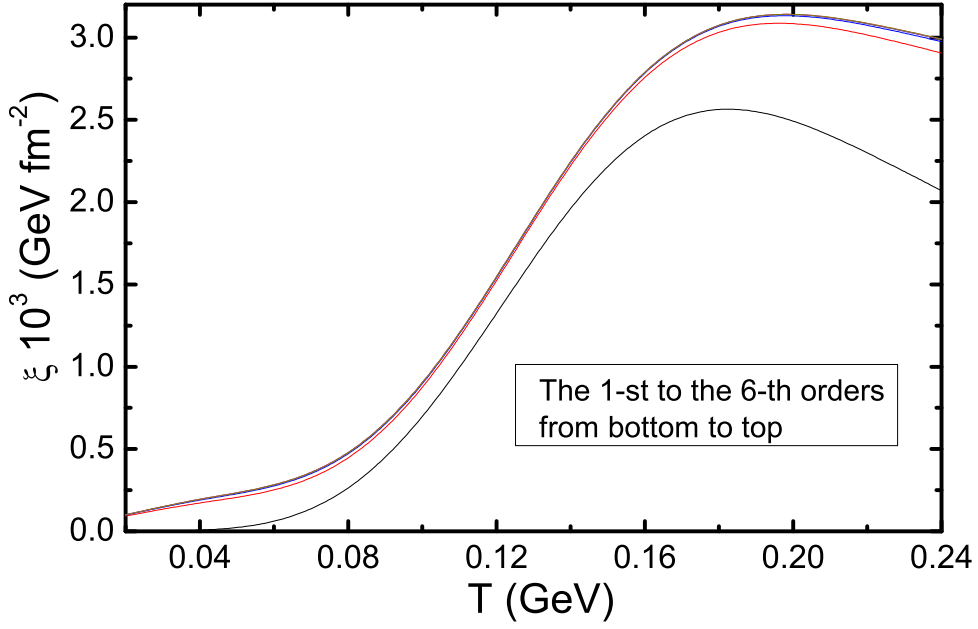,width=13cm} \caption{The bulk
viscosity as function of the temperature in the hadron gas
calculated up to the 6-th order. Chemical potentials are equal to
zero.\label{OrdersBulkAll} }
\end{center}
\end{figure}

The new particle list with charmed and bottomed particles from the THERMUS package \cite{thermus} is used. It
comprises 508 particle species including anti-particles. These are the particles, including charmed and bottomed
ones, which are more or less reliably detected \cite{Nakamura:2010zzi}. It's found that mass cut can be done on
the $3~GeV$, which results in negligible errors ($0.01\%$ or less) in all the considered quantities. The particle
list cut on the $3~GeV$ contains $462$ particle species including anti-particles.
This list is used in calculations at zero chemical potentials (throughout the paper the chemical potentials are
equal to zero if else is not stated). The results for
the shear and the bulk viscosities are shown in fig.
\ref{OrdersShearAll} and fig. \ref{OrdersBulkAll} correspondingly.
The maximal relative errors are less than $0.172\%$ for the bulk
viscosity and less than $0.444\%$ for the shear viscosity. As one
can see in the fig. \ref{CompViscosPions} the bulk viscosity of
the pion gas has maximum approximately at the temperature equal to
the half of the pion's mass. In the hadron gas this maximum shifts
towards the value ${T\approx 200~MeV}$.

\begin{figure}[h!]
\begin{center}
\epsfig{file=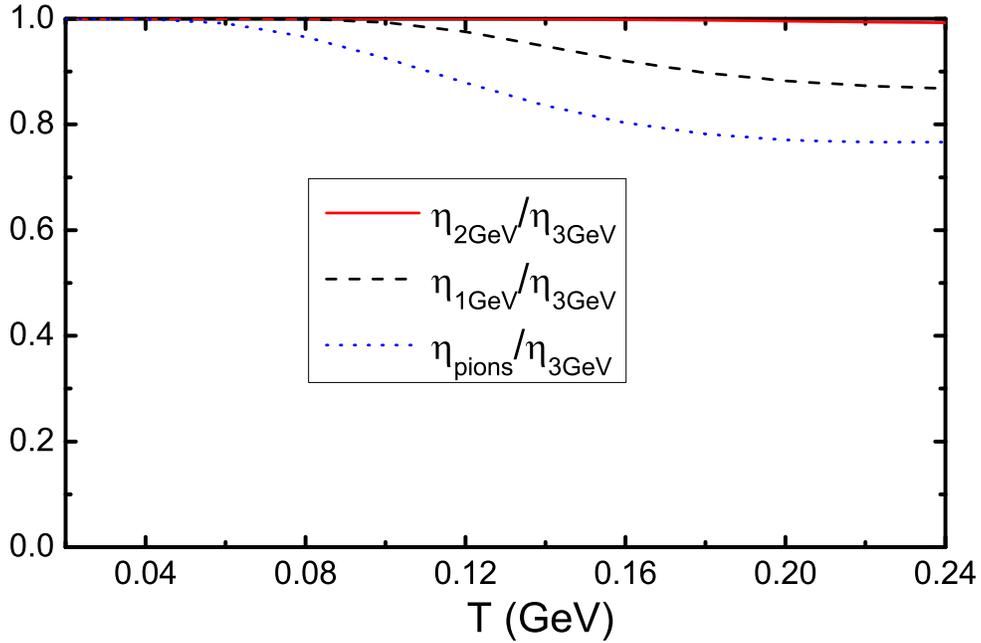,width=13cm} \caption{The ratio
of the shear viscosity $\eta_{3GeV}$ with the mass spectrum cut
on the $3~GeV$ to the shear viscosities with the mass spectrum
cut on the $2~GeV$, $1~GeV$ and on the pion's mass as functions of
the temperature. Calculations are done for zero chemical
potentials. \label{MassSpecCompShear} }
\end{center}
\end{figure}

\begin{figure}[h!]
\begin{center}
\epsfig{file=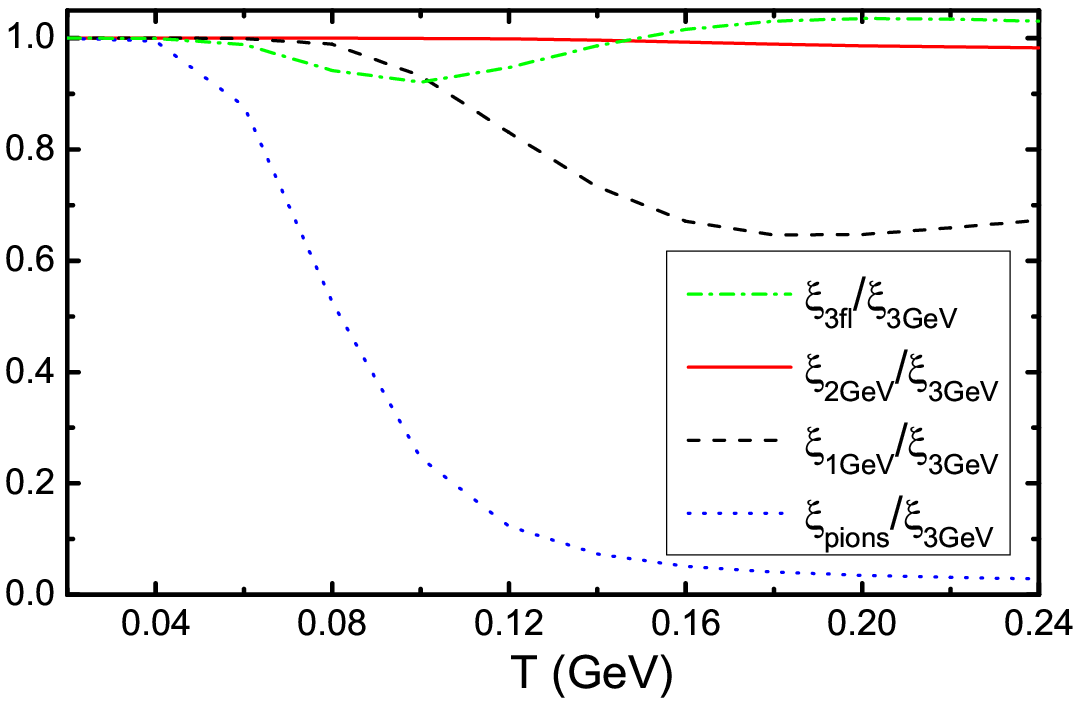,width=13cm} \caption{The ratio
of the bulk viscosity $\xi_{3GeV}$ with the mass spectrum cut on
the $3~GeV$ to the bulk viscosities with the 3 flavor hadron list, with the mass spectrum cut
on the $2~GeV$, $1~GeV$ and on the pion's mass as functions of the
temperature. Calculations are done for zero chemical potentials.
\label{MassSpecCompBulk} }
\end{center}
\end{figure}

The hadron gas mass spectrum dependence of transport coefficients is investigated. Particles in the list were
sorted over their masses and the list was cut on the $3~GeV$, $2~GeV$, $1~GeV$ and on the pion's mass. There is a
separate interest to consider also hadron list with only 3 flavors. This is because, e. g., often lattice
calculations are done with only 3 flavors. The 3 flavor list contains 358 particle species including
anti-particles. The analogical list of the UrQMD (version 1.3) \cite{urqmd} comprises 322 particle species. The
deviations due to the discrepancies in these lists are no more than $0.7\%$ in the shear viscosity and no more
than $2.1\%$ in the bulk viscosity, which can be ignored. Comparison of the shear viscosities calculated with
different hadron lists is depicted in fig. \ref{MassSpecCompShear}. As one can see the shear viscosity changes no
more than in $1.5$ times. The 3 flavor list shear viscosity is undistinguishable from the $2~GeV$ shear viscosity
(the deviations of $0.5\%$ or smaller), so that it's not shown. The bulk viscosity mass spectrum dependence is
very strong, as depicted in fig. \ref{MassSpecCompBulk}. The ratio of the bulk viscosity of the hadron gas to the
bulk viscosity of the pion gas reaches $8$ at ${T=120~MeV}$ and $13.7$ at ${T=140~MeV}$. Exclusion of charmed and
bottomed particles may result in even larger bulk viscosity values at some temperatures, as can be seen in
the fig. \ref{MassSpecCompBulk}. 

\begin{figure}[h!]
\begin{center}
\epsfig{file=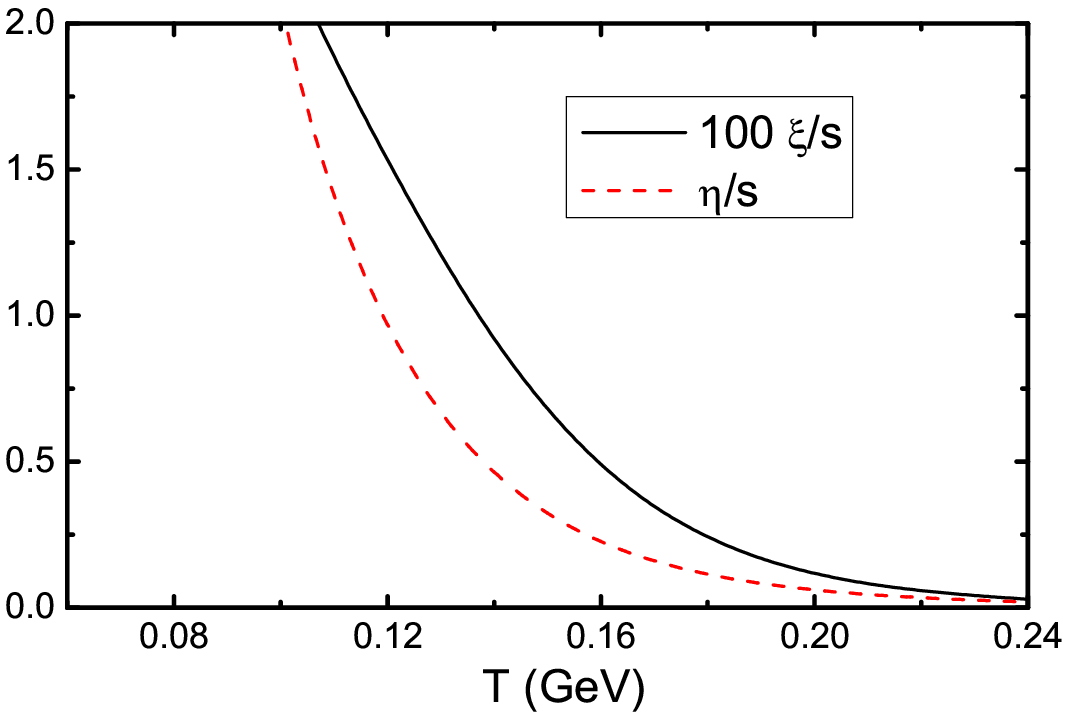,width=13cm} \caption{The ratio of the
shear viscosity to the entropy density and the ratio of the bulk
viscosity times 100 to the entropy density as functions of the
temperature in the hadron gas. Calculations are done at zero
chemical potentials. \label{EtasXis} }
\end{center}
\end{figure}

\begin{figure}[h!]
\begin{center}
\epsfig{file=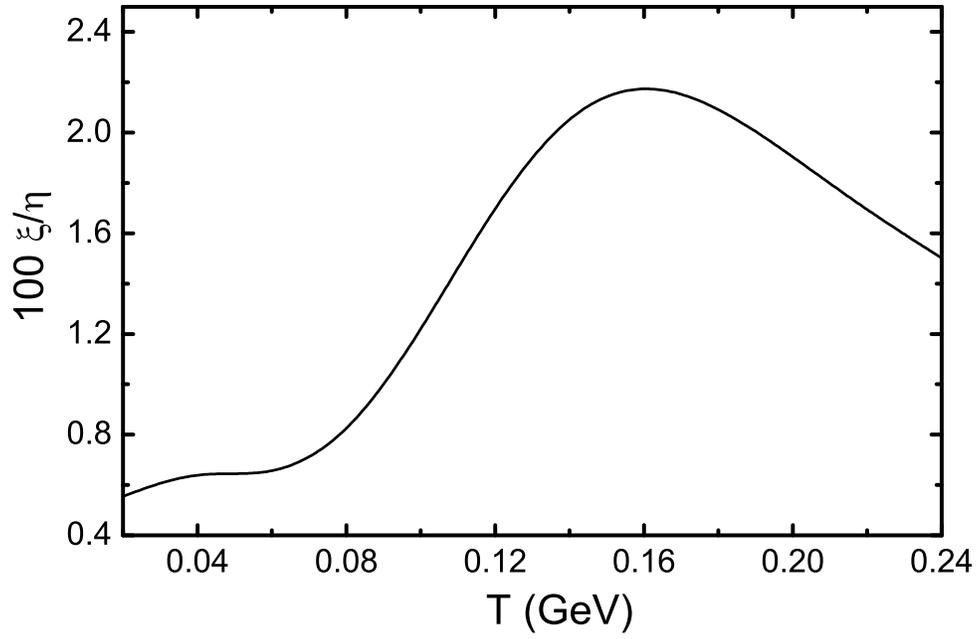,width=13cm} \caption{The ratio of the bulk
viscosity to the shear viscosity times 100 as function of the
temperature in the hadron gas. Calculations are done at zero
chemical potentials. \label{EtaXi} }
\end{center}
\end{figure}

The ratio of the shear viscosity to the entropy density $\eta/s$
and the ratio of the bulk viscosity to the entropy density $\xi/s$
in the hadron gas is shown in fig. \ref{EtasXis}. The ratio
$\xi/s$ doesn't have a maximum and is descending function of the
temperature. The entropy density is calculated by the formula
(\ref{entrden}) using the ideal gas formulas in the (\ref{ignk})
and the (\ref{assign2}). The ratio of the bulk viscosity to the
shear viscosity is shown in fig. \ref{EtaXi}.

\begin{figure}[h!]
\begin{center}
\epsfig{file=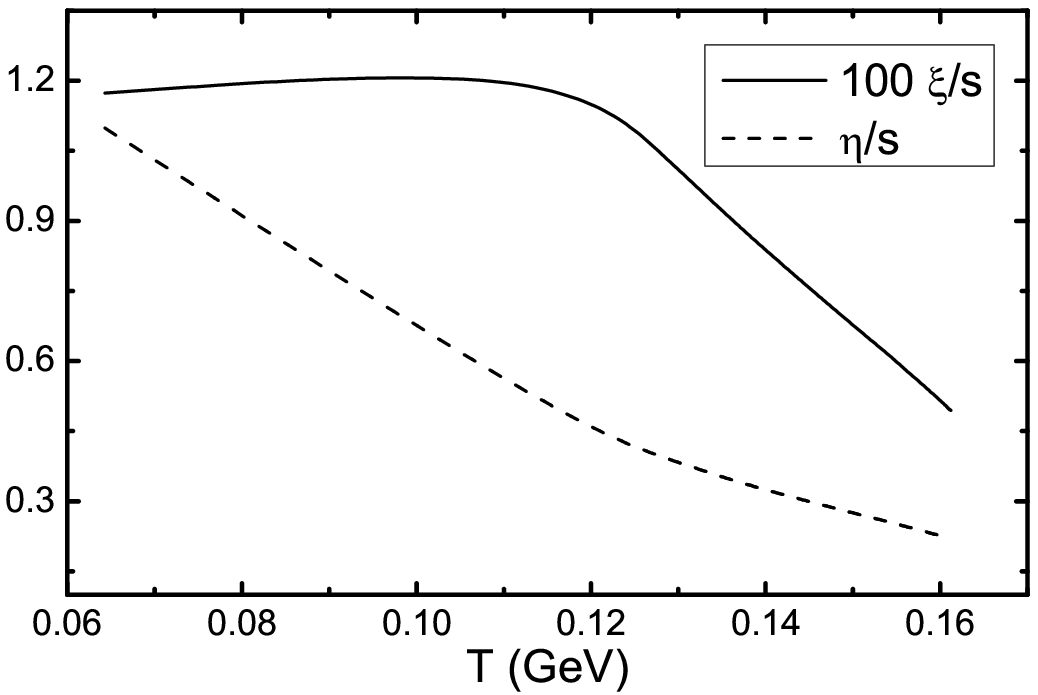,width=13cm} \caption{The shear
viscosity and the bulk viscosity times 100 as functions of the
freeze-out values of the temperature $T$. All other freeze-out
line variables' values can be found in \cite{Gorenstein:2007mw}.
\label{EtasXisFreezeout} }
\end{center}
\end{figure}

The dependence of the $\eta/s$ and the $\xi/s$ from the
temperature, calculated along the freeze-out line, is found too
and is depicted in fig. \ref{EtasXisFreezeout}. As was discussed
in Sec. \ref{condappl} calculations with large chemical potential
may contain large deviations especially for the bulk viscosity and
are rather estimating. At considered collision energies strange
particle numbers are not described well with the statistical
model. It's expected that this is because they doesn't reach
chemical equilibrium before the chemical freeze-out takes place.
After introduction of strange saturation factors $\gamma_s$
experimental data gets described well \cite{Cleymans:2005xv},
\cite{Andronic:2005yp}. These calculations were done using the old
particle list from the THERMUS package \cite{thermus} (without
charmed and bottomed particles, which comprise 358 particle
species including anti-particles). All variables' values of
freeze-out line including the strangeness saturation factor
$\gamma_s$ see in \cite{Gorenstein:2007mw}.

\section{ Analytical results }
\subsection{ The single-component gas \label{singcomsec}}
In the single-component gas with one test-function the matrix
equations can be easily solved and the shear (\ref{fineta}) and
the bulk (\ref{finxi2}) viscosities become (indexes "1" of the
particle species are omitted)
 \eq{\label{etasc}
 \eta=\frac1{10}\frac{T}{\sigma(T)}\frac{(\gamma^0)^2}{C^{00}},
 }
 \eq{\label{xisc}
 \xi=\frac{T}{\sigma(T)}\frac{(\alpha^2)^2}{A^{22}}.
 }
In this approximation the explicit closed-form (expressed through
special and elementary functions) relativistic formulas for the
bulk and the shear viscosities were obtained in \cite{anderson}
for the model with constant cross section with the ideal gas
equation of state. There the parameter ${a=2r}$. In \cite{groot}
they are written through the parameter ${\sigma=2r^2}$.
\footnote{It is the differential cross section for identical
particles. The total cross section is ${\int \frac{d\Omega}2
2r^2=4\pi r^2}$.} The results are:
 \eq{\label{etaincor}
 \eta=\frac{15}{64\pi}\frac{T}{r^2}\frac{z^2K_2^2(z)\hat h^2}
 {(5z^2+2)K_2(2z)+(3z^3+49z)K_3(2z)},
 }
 \eq{\label{xi}
 \xi=\frac1{64\pi}\frac{T}{r^2}\frac{z^2K_2^2(z)[(5-3\gamma)\hat h-3\gamma]^2}
 {2K_2(2z)+zK_3(2z)},
 }
where ${\gamma=\frac1{c_\upsilon}+1=\frac{z^2+5\hat h-\hat
h^2}{z^2+5\hat h-\hat h^2-1}}$. Though the correct result for the
shear viscosity is
 \eq{\label{eta}
 \eta=\frac{15}{64\pi}\frac{T}{r^2}\frac{z^2K_2^2(z)\hat h^2}
 {(15z^2+2)K_2(2z)+(3z^3+49z)K_3(2z)}.
 }
This result is in agreement with the result in \cite{prakash,
leeuwen}. To get the (\ref{eta}) and (\ref{xi}) the collision
brackets in the $C^{00}$ (\ref{C4ind}) and $A^{22}$ (\ref{A4ind})
can be taken from Appendix \ref{appJ} with ${z_k=z_l=z}$ and the
$\gamma^0$ and $\alpha^2$ can be taken from Appendix \ref{appA}.
In the nonrelativistic limit ${z \gg 1}$ one gets
 \eq{\label{nonrel}
 \eta=\frac{5}{64\sqrt{\pi}}\frac{T}{r^2}\sqrt{z}\left(1+\frac{25}{16}z^{-1}+...\right),
 }
 \eq{
 \xi=\frac{25}{256\sqrt{\pi}}\frac{T}{r^2}z^{-3/2}\left(1-\frac{183}{16}z^{-1}+...\right).
 }
In the ultrarelativistic limit ${z \ll 1}$ one gets
 \eq{
 \eta=\frac{3}{10\pi}\frac{T}{r^2}\left(1+\frac{1}{20}z^2+...\right),
 }
 \eq{
 \xi=\frac1{288\pi}\frac{T}{r^2}z^4\left(1+\left(\frac{49}{12}-6\ln2+6\gamma_E\right)z^2+6z^2\ln z
 +...\right).
 }
where $\gamma_E$ is the Euler's constant, ${\gamma_E\approx
0.577}$.

The perturbation of the distribution function $\varphi$
(\ref{varphi}) can be found too:
 \eq{
 \varphi=\frac1{n\sigma(T)}\left(-(A^0+A^1\tau+A^2\tau^2)\nabla_\mu U^\mu+C^0
 \overset{\circ}{\overline{\pi^\mu \pi^\nu}}
 \overset{\circ}{\overline{\nabla_\mu U_\nu}}\right),
 }
where the $C^0$ is equal to
 \eq{
 C^0=\frac{15}{64\pi}\frac{\sigma(T)}{r^2}\frac{z^2K_2^2(z)\hat h}
 {(15z^2+2)K_2(2z)+(3z^3+49z)K_3(2z)},
 }
and the $A^2$ is equal to
 \eq{
 A^2=\frac1{64\pi}\frac{\sigma(T)}{r^2}\frac{z^2K_2^2(z)[(5-3\gamma)\hat h-3\gamma]}
 {2K_2(2z)+zK_3(2z)}.
 }
The $A^0$ and the $A^1$ are used to satisfy the conditions of fit
(\ref{condfit}) and are equal to
 \eq{
 A^0=A^2\frac{a^2a^4-(a^3)^2}{\Delta_A}, \quad
 A^1=A^2\frac{a^2a^3-a^1a^4}{\Delta_A}, \quad
 \Delta_A\equiv a^1 a^3-(a^2)^2,
 }
where the $a^s$ can be found in Appendix \ref{appA}. In the
nonrelativistic limit ${z\gg 1}$ one has
 \eq{
 \varphi=\frac{5\pi e^{z-\hat\mu}}{32\sqrt2 T^3 z^2 r^2}
 \left(-(\tau^2+2z\tau-z^2)\nabla_\mu U^\mu+
 2\overset{\circ}{\overline{\pi^\mu \pi^\nu}}
 \overset{\circ}{\overline{\nabla_\mu U_\nu}}\right).
 }
In the ultrarelativistic limit $z\ll 1$ one has
 \eq{
 \varphi=\frac{\pi e^{-\hat\mu}}{480 T^3 r^2}\left(-5z^2(\tau^2+8\tau-12)\nabla_\mu
 U^\mu+36\overset{\circ}{\overline{\pi^\mu \pi^\nu}}
 \overset{\circ}{\overline{\nabla_\mu U_\nu}}\right).
 }
Note that although the shear viscosity diverges for ${T\rightarrow
\infty}$ the perturbative expansion over the gradients does not
break down because the $\varphi$ does not diverge and tends to
zero conversely.

The phenomenological formula, coming from the momentum transfer
considerations in the kinetic-molecular theory, for the shear
viscosity is ${\eta_{ph}\propto l n \langle \abs{\vec p} \rangle}$
(with the coefficient of proportionality of order 1), where
$\langle \abs{\vec p} \rangle$ is the average relativistic
momentum (\ref{avmom}), $l$ is the mean free path. It gives
correct leading $m$ and $T$ parameter dependence of the
(\ref{eta}) with quite precise coefficient. The mean free path can
be estimated as ${l\approx 1/(\sigma_{tot}n)}$ (see Appendix
\ref{appmfp}). Choosing the coefficient of proportionality to
match the nonrelativistic limit one gets \cite{Gorenstein:2007mw}
 \eq{\label{etaph}
 \eta_{ph}=\frac{5}{64\sqrt{\pi}}\frac{\sqrt{mT}}{r^2}\frac{K_{5/2}(m/T)}{K_2(m/T)}.
 }
If the bulk viscosity is expressed as ${\xi_{ph}\propto l n
\langle \abs{\vec p} \rangle}$ the coefficient of proportionality
is not of order 1. In the nonrelativistic limit it is $25/(512
\sqrt2 z^2)$ and in the ultrarelativistic limit it is $z^4/(864
\pi)$. To reproduce these asymptotical dependencies the bulk
viscosity should be proportional to the second power of the
averaged product of the source term $\hat Q$ and the $\tau$ that
is to the $(\alpha^2)^2$.

If a system has no charges, then terms proportional to the $\tau_k$ in the (\ref{Qsource}) are absent, and the
$R$ quantity gets another form. This results in quite different values of the $\alpha_k^r$. In particular for the
single-component gas in the case ${z\gg 1}$ one gets
 \eq{\label{alfrac1}
 \frac{(\alpha^2)^2|_{q_{11}=0}}{(\alpha^2)^2|_{q_{11}=1}}=\frac{4 z^4}{25}+...,
 }
and in the case ${z\ll 1}$ one gets
 \eq{\label{alfrac2}
 \frac{(\alpha^2)^2|_{q_{11}=0}}{(\alpha^2)^2|_{q_{11}=1}}=4+... .
 }
In the both cases these estimates suppose enhancement of the bulk viscosity (as can be inferred from the
(\ref{xisc})) if the number-changing processes are not negligible.

\subsection{ The binary mixture \label{binmixsec}}

The mixture of two species with masses $m_1$, $m_2$ with different
classical elastic differential constant cross sections
$\sigma^{cl}_{11}$, ${\sigma^{cl}_{12} = \sigma^{cl}_{21}}$,
$\sigma^{cl}_{22}$ is considered in this section. Using the
(\ref{fineta}) with ${n_2=0}$ and solving the matrix equation
(\ref{etamatreq}) one has for the shear viscosity:
 \eq{
 \eta=\frac{T}{10\sigma(T)}\frac1{\Delta_\eta}[(x_{1'}\gamma_1^0)^2C_{2'2'}^{00}-
 2x_{1'}x_{2'}\gamma_1^0\gamma_2^0C_{1'2'}^{00}+(x_{2'}\gamma_2^0)^2C_{1'1'}^{00}],
 }
where ${\Delta_\eta = C_{1'1'}^{00}C_{2'2'}^{00} -
(C_{1'2'}^{00})^2}$. The collision brackets for the
$C_{k'l'}^{00}$ (\ref{C4ind}) can be found in Appendix \ref{appJ}
and the $\gamma_k^0$ can be found in Appendix \ref{appA}.

In important limiting case when one mass is large ${z_2\gg 1}$
($g_2$ and $\hat \mu_2$ are finite so that ${x_{2'}\ll 1}$) and
another mass is finite one can perform asymptotic expansion of
special functions. Then one has ${x_{1'} \propto O(1)}$,
${\gamma_1^0 \propto O(1)}$, ${x_{2'} \propto
O(e^{-z_2}z_2^{3/2})}$, ${\gamma_2^0 \propto O(z_2)}$. Collisions
of light and heavy particles dominate over collisions of heavy and
heavy particles in the $C_{2'2'}^{00}$ and one has
${[\overset{\circ}{ \overline{\pi^{\mu} \pi^{\nu}} },
\overset{\circ}{ \overline{\pi_{\mu} \pi_{\nu}} }]_{21} \propto
O(z_2)}$, ${C_{2'2'}^{00} \propto O(e^{-z_2}z_2^{5/2})}$. In the
$C_{1'1'}^{00}$ collisions of light and light particles dominate
and one gets ${C_{1'1'}^{00} \propto O(1)}$. And
${[\overset{\circ}{ \overline{\pi^{\mu} \pi^{\nu}} },
\overset{\circ}{ \overline{\pi_{1\mu} \pi_{1\nu}} }]_{12} \propto
O(1)}$, ${C_{1'2'}^{00} \propto O(e^{-z_2}z_2^{3/2})}$. In the
shear viscosity the first nonvanishing contribution is the
single-component shear viscosity (\ref{eta}), where one should
take ${r^2=\sigma^{cl}_{11}}$ and ${z=z_1}$. The next correction
is
 \eq{
 \Delta \eta=z_2^{5/2}e^{-z_2}\frac{3 Tg_2 e^{z_1-\hat\mu_1+\hat\mu_2}}
 {64\sqrt{2\pi}(3+3z_1+z_1^2)g_1\sigma^{cl}_{12}}.
 }
The approximate formula \cite{Gorenstein:2007mw}
 \eq{
 \eta=\sum_k\eta_kx_k,
 }
where $\eta_k$ is given by the (\ref{eta}) or the (\ref{etaph})
with mass $m_k$ and cross section $\sigma^{cl}_{kk}$, would give
somewhat different heavy mass power dependence $O(e^{-z_2}z_2^2)$.

Using the (\ref{finxi2}) with ${n_1=1}$ and solving the matrix
equation (\ref{ximatreq2}) one has for the bulk viscosity:
 \eq{
 \xi=\frac{T}{\sigma(T)}\frac{(x_{2'}\alpha_2^1)^2}{A_{2'2'}^{11}}=
 \frac{T}{\sigma(T)}\frac{x_{1'}x_{2'}\alpha_1^1\alpha_2^1}{A_{1'2'}^{11}}.
 }
Using definition of the $A_{2'2'}^{11}$ (\ref{A4ind}) and the fact
${[\tau,\tau_1]_{kl} + [\tau,\tau]_{kl} = 0}$ (\ref{br211}) one
gets ${A_{2'2'}^{11} = x_{1'}x_{2'}[\tau,\tau]_{12}}$. Using the
(\ref{br2pos}) one gets ${[\tau,\tau]_{12} > 0}$. Then using
${x_{1'}\alpha_1^1 + x_{2'}\alpha_2^1 = 0}$, coming from the
(\ref{lhsencons}), the bulk viscosity can be rewritten as
 \eq{
 \xi=\frac{T}{\sigma(T)}\frac{x_{2'}(\alpha_2^1)^2}{x_{1'}[\tau,\tau]_{12}}
 =\frac{T}{\sigma(T)}\frac{x_{1'}(\alpha_1^1)^2}{x_{2'}[\tau,\tau]_{12}}>0.
 }
The collision bracket $[\tau,\tau]_{12}$ can be found in Appendix
\ref{appJ} and the $\alpha_k^1$ can be found in Appendix
\ref{appA}.

In the limiting case ${z_2\gg 1}$ one has ${x_{1'} \propto O(1)}$,
${x_{2'} \propto O(e^{-z_2}z_2^{3/2})}$, ${\alpha_1^1 \propto
O(e^{-z_2}z_2^{3/2})}$, ${\alpha_2^1 \propto O(1)}$, ${A_{22}^{11}
\propto A_{12}^{11} \propto O(e^{-z_2}z_2^{1/2})}$,
${[\tau,\tau]_{12} \propto O(z_2^{-1})}$. Then for the bulk
viscosity one gets
 \eq{
 \xi=e^{-z_2}z_2^{5/2}\frac{g_2 T e^{-\hat\mu_1+\hat\mu_2+z_1}[2 z_1^2-5-2\hat h_1^2
 +10 \hat h_1]^2}{128 \sqrt{2 \pi } g_1 \sigma^{cl}_{12}(z_1^2+3 z_1+3)[z_1^2-1-\hat h_1^2+5 \hat h_1]^2}+....
 }

\section{Concluding remarks}

The shear and the bulk viscosities of the hadron gas and the pion
gas were calculated at low temperatures in the model with constant
cross sections. The physics of the bulk viscosity is very
interesting. In particular it was found that in mixtures with only
elastic collisions it can strongly depend on the mass spectrum.
For instance, at temperature ${T=120~MeV}$ the bulk viscosity of
the hadron gas is larger in 8 times than the bulk viscosity of the
pion gas. Also the bulk viscosity can strongly depend on quantum
statistics corrections, equation of state and inelastic processes
which can be explained by nontrivial form of its source term.
Inclusion of inelastic processes in pion gas at ${T=140~MeV}$
results in increase of the bulk viscosity roughly in 44 times
according to comparison with results of the paper \cite{lumoore}.
It's a future task to switch off carefully inelastic processes
where they can be considered as negligible ones to perform
calculation of the bulk viscosity in the pion gas and the hadron
gas. The shear viscosity is less dependent on the mass spectrum
and on quantum statistics corrections and its source term is some
trivial function which doesn't depend on inelastic processes.

\acknowledgments

The author would like to thank to Iu. Karpenko and O. Gamayun for
comments and help in preparation of the draft.

\appendix

\section{Values of $\alpha^r_k$, $\gamma_k^r$ and $a_k^r$ \label{appA}}
Their definitions are
 \eq{\label{alphagammaadef}
 \alpha_k^r\equiv(\hat Q_k,(\tau_k)^r), \quad \gamma_k^r\equiv((\tau_k)^r
 \overset{\circ}{\overline{\pi_k^\mu\pi_k^\nu}},
 \overset{\circ}{\overline{\pi_{k\mu}\pi_{k\nu}}}), \quad
 a^s_k\equiv(1,(\tau_k)^s)_k,
 }
where the round brackets are
 \eq{
 (F,G)_k\equiv\frac1{4\pi z_k^2K_2(z_k)T^2}\int_{p_k} F(p_k)G(p_k)e^{-\tau_k}.
 }
Then one can rewrite the $a_k^s$ as
 \eq{\label{aksdef}
 a_k^s=\frac1{z_k^2K_2(z_k)}\int_{z_k}^\infty d\tau (\tau^2-z_k^2)^{1/2}\tau^s e^{-\tau}.
 }
There is recurrence relation for the $a^s_k$:
 \eq{
 a^s_k=(s+1)a^{s-1}_k+z_k^2 a^{s-2}_k-(s-2)z_k^2 a^{s-3}_k.
 }
It can be derived from the (\ref{aksdef}) written in the form
 \eq{
 a_k^s=\frac1{z_k^2K_2(z_k)}\int_{z_k}^\infty d\tau
 (\tau^2-z_k^2)^{3/2}\tau^{s-2} e^{-\tau}+z_k^2 a_k^{s-2}.
 }
Then after integration by parts the recurrence relation follow.
Some values of the $a_k^s$ are
 \eq{
 a^0_k=\frac1{z_k^2}(\hat h_k-4),
 }
 \eq{
 a^1_k=1,
 }
 \eq{
 a^2_k=\hat h_k-1,
 }
 \eq{
 a^3_k=3\hat h_k+z_k^2,
 }
 \eq{
 a^4_k=(15+z_k^2)\hat h_k+2z_k^2,
 }
 \eq{
 a^5_k=6(15+z_k^2)\hat h_k+z_k^2(15+z_k^2),
 }
 \eq{
 a^6_k=(630+45z_k^2+z_k^4)\hat h_k+5z_k^2(21+z_k^2).
 }
The $\alpha_k^r$ can be expressed through the $a^r_k$ after
integration of the (\ref{Qsource}) (or the (\ref{Qsource2}) if
only elastic collisions are considered) over momentum and using
the definition (\ref{alphagammaadef}). For systems with only
elastic collisions some values of the $\alpha_k^r$ are written
below, in agreement with \cite{groot}:
 \eq{\label{elal0}
 \alpha_k^0=0,
 }
 \eq{
 \alpha_k^1=\frac{2(c_\upsilon-9)\hat h_k+3\hat h_k^2-3z_k^2}{c_\upsilon}
 =\frac{\gamma_k-\gamma}{\gamma_k-1},
 }
 \eq{
 \alpha_k^2=2\hat h_k-3\frac{c_{\upsilon,k}}{c_\upsilon}-3\frac{\hat
 h_k+1}{c_\upsilon}=(5-3\gamma)\hat
 h_k-3\gamma_k\frac{\gamma-1}{\gamma_k-1},
 }
where the assignments $\gamma$ and $\gamma_k$ from \cite{groot}
are used. They can be expressed through the $c_\upsilon$ and the
$c_{\upsilon,k}$, defined in the (\ref{elquant}), as
 \eq{
 \gamma\equiv \frac1{c_\upsilon}+1, \quad
 \gamma_k\equiv\frac1{c_{\upsilon,k}}+1.
 }
The $\gamma_k^r$ can be rewritten as
 \eq{
 \gamma_k^r=\frac23\frac1{z_k^2K_2(z_k)}\int_{z_k}^\infty d\tau
 (\tau^2-z_k^2)^{5/2}\tau^r e^{-\tau}.
 }
Then it can be rewritten through the $a_k^r$:
 \eq{
 \gamma_k^r=\frac23(a^{r+4}-2z_k^2a^{r+2}+z_k^4a^r).
 }
Some values of the $\gamma_k^r$ are
 \eq{
 \gamma_k^0=10\hat h_k,
 }
 \eq{
 \gamma_k^1=10(6\hat h_k+z_k^2),
 }
 \eq{
 \gamma_k^2=10(7z_k^2+\hat h_k(42+z_k^2)).
 }

\section{The entropy density formula \label{appTherm}}
The Gibbs's potential is defined as
 \eq{\label{Phidef}
 \Phi(P,T)\equiv E(S,V)-ST+PV.
 }
The differential of energy is defined as
 \eq{
 dE=TdS-PdV+\sum_k\mu_kdN_k=TdS-PdV+\sum_a\mu_adN_a,
 }
where it is rewritten through the independent chemical potentials
and the particle net charges $N_a$. The differential of the $\Phi$
then reads:
 \eq{\label{dPhi}
 d\Phi=-SdT+VdP+\sum_a \mu_a dN_a.
 }
Because the $\Phi$ is function of intrinsic variables $P$, $T$ and
extrinsic $N_a$ the only possible form of it in the thermodynamic
limit is
 \eq{\label{Phi}
 \Phi=\sum_a N_a \phi_a(P,T),
 }
where $\phi_a$ are unknown functions. Then from the (\ref{dPhi})
one gets ${\frac{\p \Phi}{\p N_a} = \mu_a}$, which means that
${\phi_a = \mu_a}$. Then substituting the (\ref{Phi}) into the
(\ref{Phidef}) one gets the relation
 \eq{
 \sum_a N_a \mu_a(P,T)=E(S,V)-ST+PV.
 }
Being written for local infinitesimal volume it transforms into
the expression
 \eq{
 \sum_a n_a \mu_a=\epsilon-sT+P,
 }
from where the entropy density $s$ can be found:
 \eq{\label{entrden}
 s=\frac{\epsilon+P}{T}-\sum_a n_a \hat \mu_a.
 }

\section{Calculation of the collision brackets \label{appJ}}
The momentum parametrization and the most transformations of the
12-dimensional integrals used below are taken from \cite{groot}
(chap. XI and XIII). Let's start from some assignments. The full
momentum is
 \eq{
 P^\mu \equiv p^\mu_k+p^\mu_{1l}={p'}^\mu_k+{p'}^\mu_{1l}.
 }
The "relative" momentums before collision $Q^\mu$ and after
collision ${Q'}^\mu$ are defined as
 \eq{
 Q^\mu=\Delta_P^{\mu\nu}(p_{k\nu}-p_{1l\nu}), \quad
 {Q'}^\mu=\Delta_P^{\mu\nu}({p'}_{k\nu}-{p'}_{1l\nu}),
 }
with the assignment
 \eq{
 \Delta_P^{\mu\nu}=g^{\mu\nu}-\frac{P^\mu P^\nu}{P^2},
 }
where $P^2\equiv P^\mu P_\mu$. The covariant cosine of the
scattering angle can be expressed through the $Q^\mu$ and the
${Q'}^\mu$ as
 \eq{
 \cos\Theta=-\frac{Q \cdot Q'}{\sqrt{-Q^2}\sqrt{-{Q'}^2}},
 }
where $\cdot$ denotes convolution of 4-vectors. One also has ${Q^2
= {Q'}^2}$ and
 \eq{\label{Q2}
 Q^2=4m_k^2-(1+\alpha_{kl})^2P^2=-\left(P^2-M_{kl}^2\right)\left[1-\frac{M_{kl}^2}{P^2}
 \left(1-\frac{4\mu_{kl}}{M_{kl}}\right)\right],
 }
where
 \eq{\label{alphakl}
 M_{kl}\equiv m_k+m_l, \quad \mu_{kl}\equiv \frac{m_k
 m_l}{m_k+m_l}, \quad  \alpha_{kl}\equiv\frac{m_k^2-m_l^2}{P^2}=\mathrm{sign}(m_k-m_l)
 \sqrt{1-\frac{4\mu_{kl}}{M_{kl}}}\frac{M_{kl}^2}{P^2}.
 }
The function $\mathrm{sign}(x)$ is equal to 1, if ${x>0}$ and
equal to $-1$, if ${x<0}$. Note that not all $P^\mu$ and $Q^\mu$
are independent:
 \eq{
 P^\mu Q_\mu=0, \quad P^\mu {Q'}_\mu=0.
 }
To come from the variables ${(p_k^\mu,p_{1l}^\mu)}$ to the
variables ${(P^\mu,Q^\mu)}$ in the measure of integration first
one has to come from the ${(p_k^\mu,p_{1l}^\mu)}$ to the
${(p_k^\mu+p_{1l}^\mu,p_k^\mu-p_{1l}^\mu)}$ (the determinant is
equal to $16$) and then shift the relative momentum
$p_k^\mu-p_{1l}^\mu$ on the $\alpha_{kl}P^\mu$. Analogically for
the ${({p'}_k^\mu,{p'}_{1l}^\mu)}$ and the ${(P^\mu,{Q'}^\mu)}$.
The inverse relations for the
$p_k^\mu,p_{1l}^\mu,{p'}_k^\mu,{p'}_{1l}^\mu$ through the
$P^\mu,Q^\mu,{Q'}^\mu$ are
 \eq{
 p_k^\mu=\frac12(1+\alpha_{kl})P^\mu+\frac12{Q}^\mu,
 }
 \eq{
 p_{1l}^\mu=\frac12(1-\alpha_{kl})P^\mu-\frac12{Q}^\mu,
 }
 \eq{
 {p'}_k^\mu=\frac12(1+\alpha_{kl})P^\mu+\frac12{Q'}^\mu,
 }
 \eq{
 {p'}_{1l}^\mu=\frac12(1-\alpha_{kl})P^\mu-\frac12{Q'}^\mu.
 }

There is a need to calculate the following integrals
 \begin{eqnarray}
  \nonumber &~& J_{kl}^{(a,b,d,e,f|q,r)} \equiv \frac{\gamma_{kl}}{T^6(4\pi)^2z_k^2z_l^2K_2(z_k)K_2(z_l)
  \sigma(T)} \int_{p_k,p_{1l},{p'}_k,{p'}_{1l}} e^{-P\cdot U/T}(1+\alpha_{kl})^q\\
  &~& \times(1-\alpha_{kl})^r \left(\frac{P^2}{T^2}\right)^a
  \left(\frac{P\cdot U}{T}\right)^b\left(\frac{Q\cdot U}{T}\right)^d\left(\frac{Q'\cdot U}{T}\right)^e
  \left(\frac{-Q\cdot {Q'}}{T^2}\right)^f W_{kl}.
 \end{eqnarray}
After nontrivial transformations described in more details in
\cite{groot} one arrives at (the constant cross section
approximation is used)
 \begin{eqnarray}\label{Jint}
  \nonumber &~&J_{kl}^{(a,b,d,e,f|q,r)}=\frac{\pi(d+e+1)!!\sigma^{(d,e,f)}_{1kl}}{z_k^2z_l^2K_2(z_k)K_2(z_l)}
  \sum_{q_1=0}^q\sum_{r_1=0}^r\sum_{k_2=0}^{\frac{d+e}2+f+1}\sum_{k_3=0}^{\frac{d+e}2+f+1}
  \sum_{h=0}^{[b/2]}(z_k+z_l)^{2(q_1+r_1+k_2+k_3)} \\
  &~&\times\left(\frac{z_k-z_l}{z_k+z_l}\right)^{q_1+r_1+2k_3}
  (-1)^{r_1+k_2+k_3+h}(2h-1)!! \binom{b}{2h}\binom{q}{q_1}\binom{r}{r_1}
  \binom{\frac{d+e}2+f+1}{k_2}\\
  \nonumber &~& \times\binom{\frac{d+e}2+f+1}{k_3} I\left(2(a+f-q_1-r_1-k_2-k_3)+3,
  b+\frac{d+e}2-h+1,z_k+z_l\right),
 \end{eqnarray}
where
 \eq{
 \sigma_{1kl}^{(d,e,f)}=\frac{\sigma^{cl}_{kl}}{\sigma(T)}
 \sum_{g=0}^{\min(d,e)} \sigma^{(f,g)} K(d,e,g),
 }
where ${\sigma^{cl}_{kl} = \gamma_{kl}\sigma_{kl}}$ is the
classical elastic differential constant cross section. The
$\sigma^{(f,g)}$ is equal to nonzero value
 \eq{
 \sigma^{(f,g)}=(2g+1) \frac{f!}{(f-g)!!(f+g+1)!!},
 }
if the difference ${f-g}$ is even and ${g \leq f}$. The $K(d,e,g)$
is equal to nonzero value
 \eq{
 K(d,e,g)=\frac{d! e!}{(d-g)!! (d+g+1)!! (e-g)!! (e+g+1)!!},
 }
if ${g \leq \min(d,e)}$ and both ${d-g}$ and ${e-g}$ are even
(which also implies that ${d+e}$ is even). The ${[...]}$ denotes
the integer part. The integral $I$ is
 \eq{\label{IIntdef}
 I(r,n,x)\equiv x^{r+n+1}\int_1^\infty du u^{r+n}K_n(xu).
 }
Also there is the following frequently used combination of the $J$
integrals:
 \eq{
 {J'}_{kl}^{(a,b,d,e,f|q,r)}\equiv\sum_{u=0}^f (-1)^u\binom{f}{u}
 (2z_k)^{2(f-u)}J_{kl}^{(a+k,b,d+e,0,0|q+2u,r)}-J_{kl}^{(a,b,d,e,f|q,r)}.
 }
The first term in the difference is obtained by the replacement of
the $Q'$ on the $Q$. Using this fact the $J'$ can be rewritten in
the form
 \begin{eqnarray}\label{Jpint}
  \nonumber &~&{J'}_{kl}^{(a,b,d,e,f|q,r)}=\frac{\pi(d+e-1)!!
  \sigma^{(d,e,f)}_{kl}}{z_k^2z_l^2K_2(z_k)K_2(z_l)}
  \sum_{q_1=0}^q\sum_{r_1=0}^r\sum_{k_2=0}^{\frac{d+e}2+f+1}\sum_{k_3=0}^{\frac{d+e}2+f+1}
  \sum_{h=0}^{[b/2]}(z_k+z_l)^{2(q_1+r_1+k_2+k_3)} \\
  &~&\times\left(\frac{z_k-z_l}{z_k+z_l}\right)^{q_1+r_1+2k_3}
  (-1)^{r_1+k_2+k_3+h}(2h-1)!! \binom{b}{2h}\binom{q}{q_1}\binom{r}{r_1}
  \binom{\frac{d+e}2+f+1}{k_2}\\
  \nonumber &~& \times\binom{\frac{d+e}2+f+1}{k_3}
  I\left(2(a+f-q_1-r_1-k_2-k_3)+3, b+\frac{d+e}2-h+1,z_k+z_l\right),
 \end{eqnarray}
where
 \eq{
 \sigma^{(d,e,f)}_{kl}=\frac{\sigma^{cl}_{kl}}{\sigma(T)}(d+e+1)\left(K(d+e,0,0)
 -\sum_{g=0}^{\min(d,e)}K(d,e,g)\sigma^{(f,g)}\right).
 }
There is recurrence relation for the integral $I$ (\ref{IIntdef})
\cite{groot, luke}
 \eq{\label{recrel}
 I(r,n,x)=(r-1)(r+2n-1)I(r-2,n,x)+(r-1)x^{r+n-1}K_n(x)+x^{r+n}K_{n+1}(x).
 }
For calculations one needs only integrals $I(r,n,x)$ with positive
values of the $n$ and odd values of the $r$. If ${r \geq -2n+1}$
the $I$ integrals can be expressed through the Bessel functions
$K_n(x)$ using the (\ref{recrel}), when ${r=1}$ or ${r=-2n+1}$.
Then using the following recurrence relation for the $K_n(x)$
\cite{luke}
 \eq{\label{Krecrel}
 K_{n+1}(x)=K_{n-1}(x)+\frac{2n}{x}K_n(x),
 }
the final result can be expressed through a couple of Bessel
functions. If ${r\leq-2n-1}$ then the recurrence relation
(\ref{recrel}) blows if one tries to express the $I(r,n,x)$
through the $I(-2n+1,n,x)$. Using the (\ref{recrel}) the $I$
integrals with ${r \leq -2n-1}$ can be expressed through the
integrals $G(n,x)$
 \eq{\label{Grecrel}
 G(n,x)\equiv I(-2n-1,n,x)=x^{-n}\int_1^\infty du u^{-n-1}K_n(xu).
 }
There is recurrence relation for the $G(n,x)$:
 \eq{
 G(n,x)=-\frac1{2n}(G(n-1,x)-x^{-n} K_n(x)).
 }
It can be easily proved by integration by parts of the
(\ref{Grecrel}) and using the following relation for the $K_n(x)$
\cite{luke}
 \eq{\label{dKdx}
 \frac{\p }{\p x}K_n(x)=-\frac{n}{x} K_n(x)-K_{n-1}(x).
 }
It is found that collision integrals have the simplest form if
they are expressed through $G(n,x)$ with ${n=3}$ or ${n=2}$ and
the Bessel functions $K_3(x)$ and $K_2(x)$ or $K_2(x)$ and
$K_1(x)$. It was chosen to take ${G(x) \equiv G(3,x)}$ and
$K_3(x)$, $K_2(x)$. The $G(x)$ can be expressed through the Meijer
function \cite{meijer}
 \eq{
 G(x)=\frac{1}{32}
 G_{1,3}^{3,0}\left((x/2)^2\left|
 \begin{array}{c}
  1 \\
  -3,0,0
 \end{array}\right.
 \right).
 }
The needed scalar collision brackets can be expressed through the
$J'$ as
 \eq{
 [\tau^r,\tau^s_1]_{kl}=\frac1{2^{r+s}}\sum_{u=1}^r\sum_{\upsilon=1}^s
 (-1)^\upsilon \binom{r}{u} \binom{s}{\upsilon} {J'}_{kl}^{(0,r+s-u-\upsilon,u,
 \upsilon,0|r-u,s-\upsilon)},
 }
 \eq{
 [\tau^r,\tau^s]_{kl}=\frac1{2^{r+s}}\sum_{u=1}^r\sum_{\upsilon=1}^s
 \binom{r}{u} \binom{s}{\upsilon} {J'}_{kl}^{(0,r+s-u-\upsilon,u,
 \upsilon,0|r+s-u-\upsilon,0)},
 }
and the needed tensorial collision brackets can be expressed as
 \begin{eqnarray}
  \nonumber &~& [\tau^r\overset{\circ}{\overline{\pi^{\mu} \pi^{\nu}}},
  \tau_1^s\overset{\circ}{\overline{\pi_{1\mu} \pi_{1\nu}}}]_{kl}=
  \frac1{2^{r+s+4}}\sum_{n_1=0}^r\sum_{n_2=0}^s \binom{s}{n_2}\binom{r}{n_1}(-1)^{s-n_2}
  ({J'}_{kl}^{(2,n_1+n_2,r-n_1,s-n_2,0|2+n_1,2+n_2)} \\
  \nonumber &+&2{J'}_{kl}^{(1,n_1+n_2,r-n_1,s-n_2,1|1+n_1,1+n_2)}+{J'}_{kl}^{(0,n_1+n_2,r-n_1,s-n_2,2|n_1,n_2)}) \\
  \nonumber &-&\frac1{2^{r+s+3}}\sum_{n_1=0}^{r+1}\sum_{n_2=0}^{s+1} \binom{s+1}{n_2}\binom{r+1}{n_1}(-1)^{s+1-n_2}
  ({J'}_{kl}^{(1,n_1+n_2,r+1-n_1,s+1-n_2,0|1+n_1,1+n_2)} \\
  \nonumber &+&{J'}_{kl}^{(0,n_1+n_2,r+1-n_1,s+1-n_2,1|n_1,n_2)})+\frac23[\tau^{r+2},\tau^{s+2}_1]_{kl}
  +\frac13 z_l^2 [\tau^{r+2},\tau^{s}_1]_{kl} \\
  &+&\frac13 z_k^2 [\tau^{r},\tau^{s+2}_1]_{kl}- \frac13 z_k^2 z_l^2
  [\tau^{r},\tau^{s}_1]_{kl},
 \end{eqnarray}
 \begin{eqnarray}
  \nonumber &~& [\tau^r\overset{\circ}{\overline{\pi^{\mu} \pi^{\nu}}},
  \tau^s\overset{\circ}{\overline{\pi_{\mu} \pi_{\nu}}}]_{kl}=
  \frac1{2^{r+s+4}}\sum_{n_1=0}^r\sum_{n_2=0}^s \binom{s}{n_2}\binom{r}{n_1}
  ({J'}_{kl}^{(2,n_1+n_2,r-n_1,s-n_2,0|4+n_1+n_2,0)} \\
  \nonumber &-&2{J'}_{kl}^{(1,n_1+n_2,r-n_1,s-n_2,1|2+n_1+n_2,0)}+{J'}_{kl}^{(0,n_1+n_2,r-n_1,s-n_2,2|n_1+n_2,0)}) \\
  \nonumber &-&\frac1{2^{r+s+3}}\sum_{n_1=0}^{r+1}\sum_{n_2=0}^{s+1} \binom{s+1}{n_2}\binom{r+1}{n_1}
  ({J'}_{kl}^{(1,n_1+n_2,r+1-n_1,s+1-n_2,0|2+n_1+n_2,0)} \\
  \nonumber &-&{J'}_{kl}^{(0,n_1+n_2,r+1-n_1,s+1-n_2,1|n_1+n_2,0)})+\frac23[\tau^{r+2},\tau^{s+2}]_{kl}
  +\frac13 z_k^2 [\tau^{r+2},\tau^{s}]_{kl} \\
  &+&\frac13 z_k^2 [\tau^{r},\tau^{s+2}]_{kl}-\frac13  z_k^4[\tau^{r},\tau^{s}]_{kl}.
 \end{eqnarray}
Below some lowest orders collision brackets are presented with the
following notations (one constant cross section approximation is
used below and $\sigma^{cl}_{kl}$ are taken equal to $\sigma(T)$):
 \begin{eqnarray}
  \nonumber \widetilde K_1 &\equiv& \frac{K_3(z_k+z_l)}{K_2(z_k)K_2(z_l)},
  \quad \widetilde K_2\equiv \frac{K_2(z_k+z_l)}{K_2(z_k)K_2(z_l)},
  \quad \widetilde K_3\equiv \frac{G(z_k+z_l)}{K_2(z_k)K_2(z_l)}, \\
  Z_{kl} &\equiv& z_k+z_l, \quad z_{kl} \equiv z_k-z_l.
 \end{eqnarray}
For the scalar collision brackets one has:
 \eq{\label{br211}
 -[\tau,\tau_1]_{kl}=[\tau,\tau]_{kl}=\frac{\pi}{2z_k^2z_l^2Z_{kl}^2}
 (P_{s1}^{(1,1)}\widetilde K_1+P_{s2}^{(1,1)}\widetilde K_2
 +P_{s3}^{(1,1)}\widetilde K_3),
 }
where
 \eq{
 P_{s1}^{(1,1)}=-2 Z_{kl} (z_{kl}^4+4 z_{kl}^2 Z_{kl}^2-2 Z_{kl}^4),
 }
 \eq{
 P_{s2}^{(1,1)}=z_{kl}^4 (3 Z_{kl}^2+8)+32 z_{kl}^2 Z_{kl}^2+8 Z_{kl}^4,
 }
 \eq{
 P_{s3}^{(1,1)}=-3 z_{kl}^4 Z_{kl}^6,
 }
and
 \eq{
 [\tau,\tau_1^2]_{kl}=[\tau^2,\tau_1]_{lk}=\frac{\pi}{4z_k^2z_l^2Z_{kl}^2}
 (P_{s11}^{(1,2)}\widetilde K_1+P_{s12}^{(1,2)}\widetilde K_2
 +P_{s13}^{(1,2)}\widetilde K_3),
 }
where
 \eq{
 P_{s11}^{(1,2)}=2 Z_{kl} (z_{kl}^5 Z_{kl}+8 z_{kl}^4+16 z_{kl}^3 Z_{kl}
 +32 z_{kl}^2 Z_{kl}^2+16 z_{kl} Z_{kl}^3-40 Z_{kl}^4),
 }
 \begin{eqnarray}
  \nonumber P_{s12}^{(1,2)}&=&-z_{kl}^5 Z_{kl} (Z_{kl}^2+8)-8 z_{kl}^4 (Z_{kl}^2+8)
  -16 z_{kl}^3 Z_{kl} (Z_{kl}^2+8)\\ &+&16 z_{kl}^2 Z_{kl}^2 (Z_{kl}^2-16)
  +8 z_{kl} Z_{kl}^3 (Z_{kl}^2-16)-8 Z_{kl}^4 (Z_{kl}^2+8),
 \end{eqnarray}
 \eq{
 P_{s13}^{(1,2)}=z_{kl}^5 Z_{kl}^7,
 }
and
 \eq{
 [\tau,\tau^2]_{kl}=[\tau^2,\tau]_{kl}=\frac{\pi}{4z_k^2z_l^2Z_{kl}^2}
 (P_{s21}^{(1,2)}\widetilde K_1+P_{s22}^{(1,2)}\widetilde K_2
 +P_{s23}^{(1,2)}\widetilde K_3),
 }
where
 \eq{
 P_{s21}^{(1,2)}=2 Z_{kl} (z_{kl}^5 Z_{kl}-8 z_{kl}^4+16 z_{kl}^3 Z_{kl}
 -32 z_{kl}^2 Z_{kl}^2+16 z_{kl} Z_{kl}^3+40 Z_{kl}^4),
 }
 \begin{eqnarray}
  \nonumber P_{s22}^{(1,2)}&=&-z_{kl}^5 Z_{kl} (Z_{kl}^2+8)+
  8 z_{kl}^4 (Z_{kl}^2+8)-16 z_{kl}^3 Z_{kl} (Z_{kl}^2+8)\\
  &-&16 z_{kl}^2 Z_{kl}^2 (Z_{kl}^2-16)+8 z_{kl} Z_{kl}^3 (Z_{kl}^2-16)
  +8 Z_{kl}^4 (Z_{kl}^2+8),
 \end{eqnarray}
 \eq{
 P_{s23}^{(1,2)}=z_{kl}^5 Z_{kl}^7,
 }
and
 \eq{
 [\tau^2,\tau_1^2]_{kl}=\frac{\pi}{24z_k^2z_l^2Z_{kl}^2}
 (P_{s11}^{(2,2)}\widetilde K_1+P_{s12}^{(2,2)}\widetilde K_2
 +P_{s13}^{(2,2)}\widetilde K_3),
 }
where
 \begin{eqnarray}
  \nonumber P_{s11}^{(2,2)}&=&-2 Z_{kl} [z_{kl}^6 (Z_{kl}^2+2)+6 z_{kl}^4
  (11 Z_{kl}^2-32)-72 z_{kl}^2 Z_{kl}^2 (Z_{kl}^2+8)\\ &+&24 Z_{kl}^4 (Z_{kl}^2+96)],
 \end{eqnarray}
 \begin{eqnarray}
  \nonumber P_{s12}^{(2,2)}&=&z_{kl}^6 (Z_{kl}^4+10 Z_{kl}^2+16)-6 z_{kl}^4
  (Z_{kl}^4-56 Z_{kl}^2+256)\\ &+&144 z_{kl}^2 Z_{kl}^2 (5 Z_{kl}^2-32)
  -48 Z_{kl}^4 (13 Z_{kl}^2+32),
 \end{eqnarray}
 \eq{
 P_{s13}^{(2,2)}=-z_{kl}^4 Z_{kl}^6 [z_{kl}^2 (Z_{kl}^2-6)-6 Z_{kl}^2],
 }
and
 \eq{
 [\tau^2,\tau^2]_{kl}=\frac{\pi}{24z_k^2z_l^2Z_{kl}^2}
 (P_{s21}^{(2,2)}\widetilde K_1+P_{s22}^{(2,2)}\widetilde K_2
 +P_{s23}^{(2,2)}\widetilde K_3),
 }
where
 \begin{eqnarray}
  \nonumber P_{s21}^{(2,2)}&=&-2 Z_{kl} [z_{kl}^6 (Z_{kl}^2+2)-36 z_{kl}^5
  Z_{kl}+18 z_{kl}^4 (Z_{kl}^2+16)+96 z_{kl}^3 Z_{kl} (Z_{kl}^2-10)\\
  &+&24 z_{kl}^2 Z_{kl}^2 (Z_{kl}^2+56)-48 z_{kl} Z_{kl}^3
  (Z_{kl}^2+20)-24 Z_{kl}^4 (Z_{kl}^2+100)],
 \end{eqnarray}
 \begin{eqnarray}
  \nonumber P_{s22}^{(2,2)}&=&z_{kl}^6 (Z_{kl}^4+10 Z_{kl}^2+16)+
  12 z_{kl}^5 Z_{kl} (Z_{kl}^2-24)-6 z_{kl}^4 (Z_{kl}^4-72 Z_{kl}^2-384)\\
  \nonumber &-&192 z_{kl}^3 Z_{kl} (Z_{kl}^2+40)-48 z_{kl}^2 Z_{kl}^2 (13 Z_{kl}^2-224)
  +96 z_{kl} Z_{kl}^3 (7 Z_{kl}^2-80)\\ &+&48 Z_{kl}^4 (13 Z_{kl}^2+48),
 \end{eqnarray}
 \eq{
 P_{s23}^{(2,2)}=-z_{kl}^4 Z_{kl}^6 [z_{kl}^2 (Z_{kl}^2-6)+12 z_{kl} Z_{kl}-6 Z_{kl}^2].
 }
And for the tensor collision brackets one has:
 \eq{
 [\overset{\circ}{\overline{\pi^{\mu} \pi^{\nu}}},
 \overset{\circ}{\overline{\pi_{1\mu} \pi_{1\nu}}}]_{kl}=
 \frac{\pi}{72z_k^2z_l^2Z_{kl}^2}(P_{T11}^{(0,0)}\widetilde K_1
 +P_{T12}^{(0,0)}\widetilde K_2+P_{T13}^{(0,0)}\widetilde K_3),
 }
where
 \begin{eqnarray}
  \nonumber P_{T11}^{(0,0)}&=&-2 Z_{kl} [z_{kl}^6 (5 Z_{kl}^2-8)+
  24 z_{kl}^4 (Z_{kl}^2-16)-144 z_{kl}^2 Z_{kl}^2 (Z_{kl}^2+8)\\
  &+&48 Z_{kl}^4 (Z_{kl}^2+72)],
 \end{eqnarray}
 \begin{eqnarray}
  \nonumber P_{T12}^{(0,0)}&=&z_{kl}^6 (5 Z_{kl}^4-40 Z_{kl}^2-64)-
  24 z_{kl}^4 (5 Z_{kl}^4+8 Z_{kl}^2+128)\\&+&576 z_{kl}^2 Z_{kl}^2
  (Z_{kl}^2-16)-192 Z_{kl}^4 (5 Z_{kl}^2+16),
 \end{eqnarray}
 \eq{
 P_{T13}^{(0,0)}=-5 z_{kl}^4 Z_{kl}^6 [z_{kl}^2 (Z_{kl}^2-24)-24 Z_{kl}^2],
 }
and
 \eq{
 [\overset{\circ}{\overline{\pi^{\mu} \pi^{\nu}}},
 \overset{\circ}{\overline{\pi_{\mu} \pi_{\nu}}}]_{kl}=
 \frac{\pi}{72z_k^2z_l^2Z_{kl}^2}(P_{T21}^{(0,0)}\widetilde K_1
 +P_{T22}^{(0,0)}\widetilde K_2
 +P_{T23}^{(0,0)}\widetilde K_3),
 }
where
 \begin{eqnarray}
  \nonumber P_{T21}^{(0,0)}&=&2 Z_{kl} [z_{kl}^6 (8-5 Z_{kl}^2)+72 z_{kl}^4 (3 Z_{kl}^2-8)
  -480 z_{kl}^3 Z_{kl} (Z_{kl}^2-4) \\ &-&336 z_{kl}^2 Z_{kl}^2 (Z_{kl}^2+8)+240 z_{kl} Z_{kl}^3
  (Z_{kl}^2+8)+192 Z_{kl}^4 (Z_{kl}^2+67)],
 \end{eqnarray}
 \begin{eqnarray}
  \nonumber P_{T22}^{(0,0)}&=&z_{kl}^6 (5 Z_{kl}^4-40 Z_{kl}^2-64)+240 z_{kl}^5
  Z_{kl}^3-24 z_{kl}^4 (5 Z_{kl}^4+48 Z_{kl}^2-192)\\ \nonumber &+&1920 z_{kl}^3 Z_{kl}
  (Z_{kl}^2-8)-192 z_{kl}^2 Z_{kl}^2 (17 Z_{kl}^2-112)+1920 z_{kl} Z_{kl}^3
  (Z_{kl}^2-8)\\ &+&768 Z_{kl}^4 (5 Z_{kl}^2+6),
 \end{eqnarray}
 \eq{
 P_{T23}^{(0,0)}=-5 z_{kl}^4 Z_{kl}^6 [z_{kl}^2 (Z_{kl}^2-24)+48 z_{kl} Z_{kl}-24 Z_{kl}^2].
 }
If ${z_k=z_l}$ then the $G(x)$ function is eliminated everywhere
and collision brackets simplify considerably.

\section{ Collision rates and mean free paths \label{appmfp}}

The quantity $\frac{W_{k'l'}} {p_{k'}^0 p_{1l'}^0 {p'}_{k'}^0
{p'}_{1l'}^0} d^3{p'}_{k'} d^3{p'}_{1l'}$, which enters in the
elastic collision integral (\ref{ckelgroot}), represents the
probability of scattering per unit time times unit volume for two
particles, which had momentums $\vec p_{k'}$ and $\vec p_{1l'}$
before scattering and momentums in ranges ${(\vec {p'}_{k'}, \vec
{p'}_{k'} + d\vec {p'}_{k'})}$ and ${(\vec {p'}_{1l'}, \vec
{p'}_{1l'} + d\vec {p'}_{1l'})}$ after scattering. The quantity $
\frac{d^3p_{k'}}{ (2\pi)^3 } f_{k'}$ represents the number of
particles per unit volume with momentums in the range ${(\vec
p_{k'}, \vec p_{k'} + d\vec p_{k'})}$. The number of collisions of
particles of the $k'$-th species with particles of the $l'$-th
species per unit time per unit volume is then\footnote{It
represents some sum over all possible collisions. In the case of
the same species one factor $\gamma_{k'l'}$ just cancels double
counting in momentum states after scattering and another factor
$\gamma_{k'l'}$ also reflects the fact that scattering takes place
for ${{n_{k'}\choose 2} \approx \frac12n_{k'}^2}$ pairs of
undistinguishable particles in a given unit volume.}
 \eq{\label{totratekl}
 \widetilde R^{el}_{k'l'}\equiv g_{k'}g_{l'}\frac{\gamma_{k'l'}^2}{(2\pi)^6}\int
 \frac{d^3p_{k'}}{p_{k'}^0}\frac{d^3p_{1l'}}{p_{1l'}^0}\frac{d^3p'_{k'}}{{p'}_{k'}^0}
 \frac{d^3p'_{1l'}}{{p'}_{1l'}^0}f_{k'}^{(0)}f_{1l'}^{(0)}W_{k'l'}.
 }
To get the corresponding number of collisions of particles of the
$k'$-th species with particles of the $l'$-th species per unit
time \emph{per particle of the $k'$-th species}, $R^{el}_{k'l'}$,
one has to divide the (\ref{totratekl}) on the
$\gamma_{k'l'}n_{k'}$ (recall that ${n_{k'} \propto g_{k'}}$ by
definition), which is the number of particles of the $k'$-th
species per unit volume divided on the number of particles of the
$k'$-th species taking part in the given type of reaction (2 for
binary elastic collisions, if particles are identical, and 1
otherwise). This rate can be directly obtained averaging the
collision rate with fixed momentum $p_k$ of the $k$-th particle
species
 \eq{
 \mc R^{el}_{kl'}\equiv g_{l'}\gamma_{kl'}\int
 \frac{d^3p_{1l'}}{(2\pi)^3}d^3p'_{k}d^3p'_{1l'}f_{1l'}^{(0)}
 \frac{W_{kl'}}{p_{k}^0p_{1l'}^0{p'}_{k}^0{p'}_{1l'}^0},
 }
over the momentum with the probability distribution
$\frac{d^3p_{k}} {(2\pi)^3} \frac{f_{k}}{n_k}$ (and spin states
which is trivial):
 \eq{
 R^{el}_{k'l'}\equiv g_{k'}g_{l'}\frac{\gamma_{k'l'}}{(2\pi)^6n_{k'}}\int
 \frac{d^3p_{k'}}{p_{k'}^0} \frac{d^3p_{1l'}}{p_{1l'}^0}
 \frac{d^3p'_{k'}}{{p'}_{k'}^0}\frac{d^3p'_{1l'}}{{p'}_{1l'}^0}f_{k'}^{(0)}
 f_{1l'}^{(0)}W_{k'l'}=\frac{\widetilde R^{el}_{k'l'}}{\gamma_{k'l'}n_{k'}}.
 }
So that to get the mean rate of elastic collisions per particle of
the $k'$-th species with all particles in the system one can just
integrate the sum of the gain terms in the collision integral
(\ref{ckelgroot}) over $\frac{d^3p_k}{(2\pi)^3p_k^0n_k}$ and
average it over spin:
 \eq{
 R_{k'}^{el}\equiv \sum_{l'} R_{k'l'}^{el}.
 }
The $\widetilde R_{k'l'}^{el}$ can be expressed through the
$J_{kl}^{(0, 0, 0, 0, 0| 0, 0)}$ integral (\ref{Jint}) as
 \begin{eqnarray}\label{Rklel}
  \nonumber \widetilde R_{k'l'}^{el}&=&\gamma_{k'l'}\sigma(T)n_{k'} n_{l'}
  J_{k'l'}^{(0, 0, 0, 0, 0| 0, 0)} \\
  &=& g_{k'}g_{l'}\gamma_{k'l'}\frac{2\sigma^{cl}_{k'l'} T^6}{\pi^3}
  [(z_{k'}-z_{l'})^2 K_2(z_{k'}+z_{l'})+z_{k'} z_{l'}(z_{k'}+z_{l'})
  K_3(z_{k'}+z_{l'})],
 \end{eqnarray}
where $\sigma^{cl}_{k'l'}$ is the classical elastic differential
constant cross section of scattering of particle of the $k'$-th
species on particles of the $l'$-th species. For the case of large
temperature or when both masses are small, ${z_{k'} \ll 1}$ and
${z_{l'} \ll 1}$, one has expansion
 \eq{
 \widetilde R_{k'l'}^{el}=g_{k'}g_{l'}\gamma_{k'l'}
 \frac{4 \sigma^{cl}_{k'l'} T^6}{\pi^3}
 \left(1-\frac14(z_{k'}^2+z_{l'}^2)+...\right).
 }
For the case of small temperature or when both masses are large,
${z_{k'}\gg 1}$ and ${z_{l'}\gg 1}$, one has expansion
 \eq{
 \widetilde R_{k'l'}^{el}=g_{k'}g_{l'}\gamma_{k'l'}\frac{\sqrt{2}\sigma^{cl}_{k'l'} T^6 z_{k'} z_{l'}
 \sqrt{z_{k'}+z_{l'}} e^{-z_{k'}-z_{l'}}}{\pi^{5/2}}\left(1+\frac{8 z_{k'}^2+19 z_{k'} z_{l'}
 +8 z_{l'}^2}{8 z_{k'} z_{l'} (z_{k'}+z_{l'})}+...\right).
 }
For the case when only one mass is large, ${z_{l'}\gg 1}$, one has
somewhat different expansion
 \eq{
 \widetilde R_{k'l'}^{el}=g_{k'}g_{l'}\gamma_{k'l'}\frac{\sqrt{2} \sigma^{cl}_{k'l'} T^6 (z_{k'}+1) z_{l'}^{3/2}
 e^{-z_{k'}-z_{l'}}}{\pi^{5/2}}\left(1+\frac{4 z_{k'}^2+15 z_{k'}+15}{8 z_{k'}+8}z_{l'}^{-1}...\right).
 }
The $\sigma(T) J_{k'l'}^{(0, 0, 0, 0, 0| 0, 0)}$ in the
(\ref{Rklel}) can be replaced in the limit of high temperatures
with $4\pi \sigma^{cl}_{k'l'} \langle \abs{\vec v_{k'}}\rangle$
and in the limit of low temperatures with $4\pi \sigma^{cl}_{k'l'}
\langle \abs{\vec v_{k'}}\rangle \sqrt{1+m_{k'}/m_{l'}}= 4\pi
\sigma^{cl}_{k'l'} \langle \abs{\vec v_{rel,k'l'}}\rangle$, where
$\langle \abs{\vec v_{k'}}\rangle$ is the mean modulus of particle
velocity of the $k'$-th species
 \eq{\label{avvel}
 \langle \abs{\vec v_{k'}}\rangle=\frac{\int d^3p_{k'} \frac{|\vec p_{k'}|}{p_{k'}^0}
 f^{(0)}_{k'}(p_{k'})}{\int d^3p_{k'} f^{(0)}_{k'}(p_{k'})}
 =\frac{2 e^{-z_{k'}}(1+z_{k'})}{z_{k'}^2K_2(z_{k'})}
 =\sqrt{\frac{8}{\pi z_{k'}}}\frac{K_{3/2}(z_{k'})}{K_{2}(z_{k'})}.
 }
and $\langle \abs{\vec v_{rel,k'l'}}\rangle$ is the mean modulus
of the relative velocity, which coincides with the $\langle
\abs{\vec v_{k'}}\rangle$ for high temperatures. Then the
resultant collision rate $R^{el}_{k'}$ would reproduce simple
nonrelativistic collision rates know in the kinetic-molecular
theory. To get the (approximate) mean free time one has just to
invert the $R^{el}_{k'}$
 \eq{
 t^{el}_{k'}=\frac{1}{R^{el}_{k'}}.
 }
The (approximate) mean free path $l^{el}_{k'}$ can be obtained
after multiplication of it on the $\langle \abs{\vec
v_{k'}}\rangle$
 \eq{\label{lel}
 l^{el}_{k'}=\frac{\langle \abs{\vec v_{k'}}\rangle}{R^{el}_{k'}}.
 }
For the single-component gas one gets
 \eq{\label{mfpsc}
 l^{el}_{1'}=\frac{\langle \abs{\vec v_{1'}}\rangle}{R^{el}_{1'1'}}=
 \frac{\pi e^{-z_{1}} (z_{1}+1)}{g_{1}4 \sigma^{cl}_{11} T^3 z_{1}^3 K_3(2 z_{1})}.
 }
The nonrelativistic limit of the (\ref{mfpsc}) with ${g_1=1}$
coincides with the same limit of the formula
 \eq{
 l^{el}_{1}=\frac{\langle \abs{\vec v_1}\rangle}{4\pi\sigma^{cl}_{11}n_1
 \langle \abs{\vec v_{rel}}\rangle}=\frac{1}{4\pi\sigma^{cl}_{11}n_1\sqrt2},
 }
which is the mean free path formula coming from the
nonrelativistic kinetic-molecular theory obtained by Maxwell. The
ultrarelativistic limit of the (\ref{mfpsc}) with ${g_1=1}$
coincides with the same limit of the formula
 \eq{
 l_1^{el}=\frac{1}{4\pi\sigma^{cl}_{11}n_1}.
 }

Analogically one can introduce inelastic rates $R^{inel}_k$ of any
inelastic processes to occur for the $k'$-th particles species.
Then the mean free time $t^{inel}_{k'}$, in what any inelastic
process for particles of the $k'$-th species occurs, can be
introduced as
 \eq{\label{tkinel}
 t^{inel}_{k'}=\frac{1}{R^{inel}_{k'}}.
 }
The mean free path for particles of the $k'$-th species is
obtained through the rate ${R^{el}_{k'} + R^{inel}_{k'}}$ and can
be written as
 \eq{\label{mfp}
 l_{k'}=\frac{\langle \abs{\vec v_{k'}}\rangle}{R^{el}_{k'}+R^{inel}_{k'}}.
 }

\end{document}